# A coupled finite-volume solver for numerical simulation of electrically-driven flows

F. Pimenta, M.A. Alves[*]

CEFT, Departamento de Engenharia Química, Faculdade de Engenharia da Universidade do Porto, Rua Dr. Roberto Frias, 4200-465 Porto, Portugal

**Abstract**

The accuracy and stability of implicit CFD codes are frequently impaired by the decoupling between variables, which can ultimately lead to numerical divergence. Coupled solvers, which solve all the governing equations simultaneously, have the potential to fix this problem. In this work, we report the implementation of coupled solvers for transient and steady-state electrically-driven flow simulations in the finite-volumes framework. The numerical method, developed in OpenFOAM®, is generic for Newtonian and viscoelastic fluids and is formulated for the Poisson-Nernst-Planck and Poisson-Boltzmann models. The resulting coupled systems of equations are solved efficiently with PETSc library. The performance of the coupled solvers is assessed in two test cases: induced-charge electroosmosis of a Newtonian fluid around a cylinder; electroosmotic flow of a PTT viscoelastic fluid in a contraction/expansion microchannel. The coupled solvers are more accurate in transient simulations and allow the use of larger time-steps without numerical divergence. For steady-state simulations, the coupled solvers converge in fewer iterations than segregated solvers. Although coupled solvers are much slower in a *per* time-step basis, the overall speedup factor obtained in this study reached a maximum value of ~100, where the highest factors have been obtained with semi-coupled solvers, which drop some coupling terms between equations. While further research is needed to improve the efficiency of the matrix solving stage, coupled solvers are already superior to segregated solvers in a number of cases.

[*] Corresponding author.
Email addresses: fm.ppimenta@gmail.com (F. Pimenta), mmalves@fe.up.pt (M.A. Alves).

# 1. Introduction

The discretization of the governing equations in computational fluid dynamics (CFD) methods such as finite-differences, finite-elements and finite-volumes results in a system of equations for each unknown variable. For example, the discretization of the momentum and continuity equations in a 3D incompressible flow gives rise to a system of equations for $p$, $u_x$, $u_y$ and $u_z$. In the finite-volumes method (FVM), it is common practice to solve each system of equations individually and in sequence, which is known as the segregated solution method. Under this approach, there is only one unknown field per equation, and special methods are needed to ensure the coupling between variables and avoid numerical divergence. Predictor-corrector methods that couple velocity and pressure (primitive variables), such as projection [1] and pressure-correction based methods (SIMPLE-type and PISO algorithms) [2], are illustrative of such methods. However, because these techniques are generally not fully-implicit and make use of approximations (e.g. the approximation of the inverse of momentum equation in projection methods [3] and the drop of the neighbor correction terms in SIMPLE algorithm [2]), they suffer from stability and accuracy problems. Here stability is used as a synonym of numerical robustness, and, in particular, as the capability of a solver to reach the expected solution (free of numerical artifacts) with minimum under-relaxation and/or the highest time-step possible, without numerical divergence.

For example, the contradictory existence of a limiting time-step (or Courant number) in a fully-implicit finite-volume code (classification based on the discretization of time-derivatives) is frequently the visible consequence of the loss of accuracy and stability of pressure-correction based methods (e.g. [4]). The presence of numerical artifacts in transient low Reynolds number flows is yet another visible deficiency of some projection methods [3]. In addition, the poor scaling behavior of SIMPLE-type algorithms with mesh refinement can be also mentioned, as shown for example by Darwish et al. [5], which was attributed in part to the use of under-relaxation [6], and that is responsible for performance deterioration.

The stability and accuracy issues arising from the decoupling between variables have been tackled in different ways. Considering the pressure-velocity coupling as a reference example, vorticity-based formulations [7] of the Navier-Stokes equations are an effective way of circumventing that problem, since the pressure variable is removed from the set of unknowns. This method relies on re-writing the governing equations in terms of derived variables and explores the relationship between such variables and mathematical operators. However, vorticity-based formulations loose generality, since extending them to 3D flows and defining appropriate boundary conditions is not straightforward. Another method, the one explored in this work, is solving the set of governing equations coupled [6, 8-10]. In contrast with the segregated solution method, coupled solvers assemble all the governing equations into a single matrix, and some terms ensure the implicit coupling between equations. The coupled solution method operates at the matrix level, thus it can be theoretically applied to any set of equations, but its efficacy in strengthening the numerical stability depends on the existence of linear terms that can be discretized implicitly and used to couple the equations. Coupled solvers have been used for a long time, but segregated solvers have a higher acceptance among finite-volumes due to its reasonable performance, low memory usage and reduced computational cost per time-step. Recent examples on the use of coupled solvers in the finite-volume framework include its application to compressible [9, 11, 12] and incompressible [6, 8, 13] flows of Newtonian fluids, and to steady laminar viscoelastic fluid flows [10].

For electrically-driven flows (EDF), in addition to the problematic pressure-velocity coupling, there is also the coupling between the electric potential and species concentration that should be considered in the Poisson-Nernst-Planck (PNP) system of equations [14-16]. It is a common procedure to apply a segregated solution method to solve these equations (e.g. [15, 17, 18]), without any implicit coupling between variables. Stability and accuracy can be improved by inner iterating multiple times the PNP system of equations within the same time-step [14, 15]. This simple method has been used successfully in a variety of EDF simulations [15, 17, 18]. A semi-implicit coupling can also be imposed in the species transport equation by proper manipulation of the electromigration term, making use of the Poisson equation for the electric potential [14]. However, we observed that both methods failed in some EDF, as in the simulation of chaotic induced-charge electroosmosis [19], addressed later in this work. The failure suggested the need for a stronger coupling between variables, which has motivated the present study.

This work addresses the implementation of a coupled solution method for EDF in order to improve the numerical stability and temporal accuracy of the segregated solver that we previously implemented in OpenFOAM® [15], and that has been incorporated in rheoTool [20]. The coupling algorithm is developed considering Newtonian fluids and also more complex viscoelastic fluid models, which can show unusual behavior in EDF [21-24]. In addition, we also formulate and test the solver with the Poisson-Boltzmann (PB) model, which is often used in replacement of the PNP system of equations [15, 24, 25]. Throughout this work, the coupled solvers are systematically compared against the segregated solver in two test cases involving both transient and steady-state flows. These test cases are the induced-charge electroosmosis (ICEO) of a Newtonian fluid around a metallic cylinder and the electroosmotic flow of a PTT fluid in a contraction/expansion microchannel. To the best of our knowledge, the performance of coupled solvers compared to segregated solvers in EDF was not assessed before in the context of finite-volumes, neither for Newtonian nor viscoelastic fluids. The application of coupled solvers to transient flows was also seldom investigated. Therefore, this work can provide important contributions to these fields of research. Note, however, that the use of coupled solvers for EDF simulations has already been reported in previous studies (e.g. [16, 26]), but within a different context and exclusively for Newtonian fluids and the PNP model.

The remainder of this work is organized as follows. In Section 2 we present the governing equations for EDF of Newtonian and viscoelastic fluids. In Section 3, the numerical implementation of the coupled solvers is described, which includes the matrix organization of the coupled system of equations, the implicit discretization of the coupling terms and the interface specifically built to solve the resulting coupled system of equations. Section 4 provides a brief description of the hardware used in this work and the results obtained are presented in Section 5. A brief discussion of the results is presented in Section 6, and the concluding remarks of Section 7 close this work.

## 2. Governing equations

Consider, as a general case, the electrically-driven, laminar, isothermal flow of an incompressible, viscoelastic fluid. The mass conservation and momentum balance can be expressed through Eqs. (1) and (2), respectively [15, 27],

$$\nabla \cdot \mathbf{u} = 0 \tag{1}$$

$$\rho \left( \frac{\partial \mathbf{u}}{\partial t} + \mathbf{u} \cdot \nabla \mathbf{u} \right) = -\nabla p + \eta_S \nabla^2 \mathbf{u} + \nabla \cdot \boldsymbol{\tau} + \mathbf{f}_E \tag{2}$$

where $\mathbf{u}$ is the velocity vector, $t$ is the time, $p$ is the pressure, $\boldsymbol{\tau}$ is the extra-stresses tensor, $\mathbf{f}_E$ represents the electric force per unit volume, $\rho$ is the fluid density and $\eta_S$ is the solvent viscosity. In order to increase the numerical stability of viscoelastic fluid flow simulations, the both-sides-diffusion technique [28] is employed to solve Eq. (2), which consists in the addition of the explicit/implicit term $-\eta_P \nabla^2 \mathbf{u}$ to both sides of the momentum equation ($\eta_P$ is the polymeric viscosity). Note that this technique is different from the stress-velocity coupling algorithm described in [27] for an Oldroyd-B fluid, which is itself equivalent to the so-called improved both-sides-diffusion technique proposed in [29]. If the fluid is Newtonian, then $\eta_S = \eta$, $\eta_P = 0$ and $\boldsymbol{\tau} = \mathbf{0}$.

Several constitutive equations are available to model the viscoelastic properties of complex fluids. In this work, the simplified linear Phan-Thien-Tanner (PTT) viscoelastic model [30] was selected to illustrate the application of coupled solvers to complex fluids. This model captures both the shear-thinning and elastic behavior of several fluids. Note, however, that the coupling strategy reported here can be easily extended to other constitutive equations sharing the same coupling terms. The constitutive equation of the simplified linear PTT model can be expressed as [30],

$$\left[1 + \frac{\varepsilon\lambda}{\eta_P}\text{tr}(\boldsymbol{\tau})\right]\boldsymbol{\tau} + \lambda\overset{\nabla}{\boldsymbol{\tau}} = \eta_P\left(\nabla\mathbf{u} + \nabla\mathbf{u}^T\right) \qquad (3)$$

where $\lambda$ is the relaxation time of the fluid, $\varepsilon$ is the extensibility parameter controlling the degree of shear-thinning and $\overset{\nabla}{\boldsymbol{\tau}} = \frac{\partial \boldsymbol{\tau}}{\partial t} + \mathbf{u}\cdot\nabla\boldsymbol{\tau} - \boldsymbol{\tau}\cdot\nabla\mathbf{u} - \nabla\mathbf{u}^T\cdot\boldsymbol{\tau}$ represents the upper-convected time derivative.

In the presence of an electric field and charged species, and neglecting magnetic effects and variations in the electric permittivity of the fluid, the electric body-force can be expressed as [15]

$$\mathbf{f}_E = -\rho_E\nabla\Psi \qquad (4)$$

where $\rho_E$ is the charge density and $\Psi$ represents the electric potential. Denoting Faraday's constant as $F$, $z_i$ and $c_i$ representing the charge and volumetric concentration of the $m$ charged species present in the fluid, then the charge density is

$$\rho_E = F\sum_{i=1}^{m} z_i c_i \qquad (5)$$

and the transport equation for each species is governed by

$$\frac{\partial c_i}{\partial t} + \mathbf{u}\cdot\nabla c_i = \nabla\cdot(D_i\nabla c_i) + \nabla\cdot\left[\left(D_i\frac{ez_i}{kT}\nabla\Psi\right)c_i\right] \qquad (6)$$

where $D_i$ represents the diffusivity of species $i$, $e$ is the elementary electric charge, $k$ is Boltzmann's constant and $T$ is the absolute temperature. The Poisson equation determining the distribution of electric potential is given by

$$\nabla\cdot(\varepsilon_r\varepsilon_0\nabla\Psi) = -\rho_E \qquad (7)$$

where $\varepsilon_0$ is the electric permittivity of vacuum and $\varepsilon_r$ represents the electric permittivity of the fluid relative to vacuum. Eqs. (5)–(7) form the PNP system of equations.

In some situations, the PNP equations can be simplified to the PB model, which assumes that the ions are in thermodynamic equilibrium and follow a Boltzmann distribution [25]. The Poisson-Boltzmann model tends to reduce the numerical stiffness of the problem since the species concentration becomes algebraically related to the electric potential, which remains as the unique unknown [25]. According to this model, the electric charge density is given by [15]

$$\rho_E = F \sum_{i=1}^{m} z_i c_{i,0} \exp\left(-\frac{ez_i}{kT}\psi\right) \tag{8}$$

and Eq. (7) reduces to

$$\nabla \cdot (\varepsilon_r \varepsilon_0 \nabla \psi) = -F \sum_{i=1}^{m} z_i c_{i,0} \exp\left(-\frac{ez_i}{kT}\psi\right) \tag{9}$$

We should note that $\psi$ in Eqs. (8) and (9) represents the intrinsic electric potential, associated with the electric double-layer (EDL), and co-exists with the externally imposed electric potential ($\varphi$). The form taken by Eq. (9) assumes that the species concentration is $c_{i,0}$ where the intrinsic potential is null ($\psi = 0$), typically in the bulk of a solution, far from charged surfaces (outside the EDL). When the total electric potential is split in this way ($\Psi = \psi + \varphi$), a Laplace equation needs to be solved for the imposed electrical potential to compute its spatial distribution (see [15] for more details),

$$\nabla \cdot (\varepsilon_r \varepsilon_0 \nabla \varphi) = 0 \tag{10}$$

In electrically-driven flows, and more specifically in electroosmosis, the flow is sustained by the movement of a thin layer of charged ions next to a charged surface, the so-called electric-double layer [31]. The Debye length ($\lambda_D$) is an important parameter providing an estimate of the EDL size,

$$\lambda_D = \sqrt{\frac{\varepsilon_r \varepsilon_0 kT}{eF \sum_{i=1}^{m} z_i^2 c_{i,0}}} \tag{11}$$

## 3. Numerical method

The numerical implementation of the equations presented in the previous section has been discussed in detail in [15, 27] for a segregated solution method. These equations were discretized in the finite-volume framework of OpenFOAM® and are available in rheoTool [20]. In this work, we limit our discussion to the modifications introduced by coupled solvers, which includes the implicit discretization of coupling terms and the solution of the resulting coupled system of equations.

### 3.1. Coupled solver

For the general case of the EDF of a viscoelastic fluid using the PNP model (assuming two ionic species denoted as '+' and '-', without loss of generality), the final coupled system of equations to be solved is represented by Eq. (12).

$$\begin{array}{l}\text{Momentum equation} \\ \text{Continuity equation} \\ \text{Constitutive equation} \\ \text{Poisson equation} \\ \text{Species transport equation}\end{array}\left\{\begin{bmatrix} a_{ux}^{ux} & & & a_{ux}^{p} & a_{ux}^{\tau xx} & a_{ux}^{\tau xy} & a_{ux}^{\tau xz} & a_{ux}^{\tau yy} & a_{ux}^{\tau yz} & a_{ux}^{\tau zz} & a_{ux}^{\Psi} & & & \\ & a_{uy}^{uy} & & a_{uy}^{p} & a_{uy}^{\tau xx} & a_{uy}^{\tau xy} & a_{uy}^{\tau xz} & a_{uy}^{\tau yy} & a_{uy}^{\tau yz} & a_{uy}^{\tau zz} & a_{uy}^{\Psi} & & & \\ & & a_{uz}^{uz} & a_{uz}^{p} & a_{uz}^{\tau xx} & a_{uz}^{\tau xy} & a_{uz}^{\tau xz} & a_{uz}^{\tau yy} & a_{uz}^{\tau yz} & a_{uz}^{\tau zz} & a_{uz}^{\Psi} & & & \\ a_{p}^{ux} & a_{p}^{uy} & a_{p}^{uz} & a_{p}^{p} & & & & & & & & & & \\ a_{\tau xx}^{ux} & a_{\tau xx}^{uy} & a_{\tau xx}^{uz} & & a_{\tau xx}^{\tau xx} & & & & & & & & & \\ a_{\tau xy}^{ux} & a_{\tau xy}^{uy} & a_{\tau xy}^{uz} & & & a_{\tau xy}^{\tau xy} & & & & & & & & \\ a_{\tau xz}^{ux} & a_{\tau xz}^{uy} & a_{\tau xz}^{uz} & & & & a_{\tau xz}^{\tau xz} & & & & & & & \\ a_{\tau yy}^{ux} & a_{\tau yy}^{uy} & a_{\tau yy}^{uz} & & & & & a_{\tau yy}^{\tau yy} & & & & & & \\ a_{\tau yz}^{ux} & a_{\tau yz}^{uy} & a_{\tau yz}^{uz} & & & & & & a_{\tau yz}^{\tau yz} & & & & & \\ a_{\tau zz}^{ux} & a_{\tau zz}^{uy} & a_{\tau zz}^{uz} & & & & & & & a_{\tau zz}^{\tau zz} & & & & \\ & & & & & & & & & & a_{\Psi}^{\Psi} & a_{\Psi}^{c+} & a_{\Psi}^{c-} \\ & & & & & & & & & & a_{c+}^{\Psi} & a_{c+}^{c+} & \\ & & & & & & & & & & a_{c-}^{\Psi} & & a_{c-}^{c-} \end{bmatrix} \begin{bmatrix} u_x \\ u_y \\ u_z \\ p \\ \tau_{xx} \\ \tau_{xy} \\ \tau_{xz} \\ \tau_{yy} \\ \tau_{yz} \\ \tau_{zz} \\ \Psi \\ c_+ \\ c_- \end{bmatrix} = \begin{bmatrix} b_{ux} \\ b_{uy} \\ b_{uz} \\ b_p \\ b_{\tau xx} \\ b_{\tau xy} \\ b_{\tau xz} \\ b_{\tau yy} \\ b_{\tau yz} \\ b_{\tau zz} \\ b_{\Psi} \\ b_{c+} \\ b_{c-} \end{bmatrix}\right.$$

(12)

In this equation, each block matrix coefficient $a_{\beta}^{\alpha}$ represents the contribution from variable $\alpha$ to the equation written for variable $\beta$ and is itself an $N \times N$ sparse matrix, where $N$ is the number of cells of the computational domain. For a 3D case, the block matrix in Eq. (12) is sparse and has a total of $13N$ rows and $13N$ columns. The square matrix represented by each coefficient $a_{\beta}^{\alpha}$ can be further decomposed as $a_{\beta}^{\alpha} = d_{\beta}^{\alpha} + o_{\beta}^{\alpha}$, where matrix $d_{\beta}^{\alpha}$ contains all the diagonal elements of $a_{\beta}^{\alpha}$ and matrix $o_{\beta}^{\alpha}$ contains all the off-diagonal elements of $a_{\beta}^{\alpha}$.

The off-diagonal blocks in the matrix of Eq. (12) arise from the implicit discretization of the coupling terms in the governing equations. These terms ensure the stability of coupled solvers and are the main difference compared to segregated solvers. The diagonal blocks result from the discretization of the remaining terms of the governing equations. For the sake of conciseness, in this work we will only discuss the origin of the off-diagonal blocks, since the diagonal blocks have been addressed in previous works [15, 27], as their discretization is common to segregated solvers. For example, the diagonal coefficients $a_{ux}^{ux}, a_{uy}^{uy}$ and $a_{uz}^{uz}$ (Eq. 12) arise from the time-derivative, convection and diffusion operators in the momentum equation (Eq. 2) and their discretization is the same both in the coupled and segregated solvers. On the other hand, the pressure gradient, extra-stresses divergence and electric force terms in that same equation (Eq. 2) are discretized implicitly in the coupled solver, being incorporated as off-diagonal coefficients (coefficients in rows 1–3 and columns 4–11 of Eq. 12), but are discretized explicitly and sent to the source terms vector in the segregated solver.

Note that the terms of the governing equations that we classify as *coupling terms* exist in both segregated and coupled solvers, and the same discretization schemes are applied in either case, although this is not mandatory. For example, in this work the volume integration of the gradient terms was discretized using the Green-Gauss theorem. This is convenient for coupled solvers because the term becomes linear in $p$, but merely optional for segregated solvers (a least-squares reconstruction could be used instead, for example). Therefore, since the discretization schemes are common, the main difference regarding the *coupling terms* lies in the contribution to the matrix of coefficients and source vector in each case, i.e. their implicit discretization in coupled solvers and explicit discretization in segregated solvers.

For the Poisson-Boltzmann model, Eq. (12) only has the coefficients concerning **u**, $p$ and **τ**, since Eqs. (9) and (10), governing the electric potential, are solved individually, and the species concentration is not an unknown in that model.

The numerical implementation of the boundary conditions used in this work is similar for both coupled and segregated solvers. Such boundary conditions are either implicit or explicit in the own variable [15, 27], but there is no implicit coupling (at the matrix level) between variables.

### 3.1.1. Coupling between pressure and velocity

The implicit coupling between pressure and velocity, in a coupled solver, has been already addressed in other works (e.g. [6, 8]). The pressure couples to velocity in the momentum equation (Eq. 2), through term $-\nabla p$, which is discretized implicitly using the Green-Gauss theorem. Consider a cell **P** with an arbitrary number of faces, such that each interior face is shared between cell **P** and a neighbor cell **N** (usually different for each face). In this case, the implicit discretization of the pressure gradient in cell **P** results in [6]

$$-\int_{V_P} \nabla p \, dV_P = -\sum_f \mathbf{S}_f p_f = -\sum_f \begin{bmatrix} S_x \\ S_y \\ S_z \end{bmatrix} (p_P w_P + p_N w_N) \tag{13}$$

where

$$\begin{cases} w_P = \dfrac{s_{Nf}}{s_{NP}} = \dfrac{|\mathbf{S}_f \cdot (\mathbf{x}_f - \mathbf{x}_N)|}{|\mathbf{S}_f \cdot (\mathbf{x}_f - \mathbf{x}_N)| + |\mathbf{S}_f \cdot (\mathbf{x}_f - \mathbf{x}_P)|} \\ w_N = \dfrac{s_{Pf}}{s_{NP}} = \dfrac{|\mathbf{S}_f \cdot (\mathbf{x}_f - \mathbf{x}_P)|}{|\mathbf{S}_f \cdot (\mathbf{x}_f - \mathbf{x}_N)| + |\mathbf{S}_f \cdot (\mathbf{x}_f - \mathbf{x}_P)|} = 1 - w_P \end{cases} \tag{14}$$

are geometric weighting factors, dependent on the position vector of cell **P** ($\mathbf{x}_P$), cell **N** ($\mathbf{x}_N$) and face $f$ ($\mathbf{x}_f$). In addition, $\mathbf{S}_f = \mathbf{n}_f A_f$ corresponds to the face-area vector, with $\mathbf{n}_f$ the unitary face-normal vector and $A_f$ the area of the face. For static grids, these factors are constant during the whole simulation. For the ease of notation, in Eqs. (13) and (14), as well as in the remaining equations throughout this work, subscript $f$ is dropped from the terms dependent on each neighbor cell **N** (for example, $S_x, S_y, S_z, w_P, p_N$ and $w_N$ in Eq. (13) should be formally written as $S_{x,f}, S_{y,f}, S_{z,f}, w_{P,f}, p_{N,f}$ and $w_{N,f}$, respectively). The contribution of term $-\nabla p$ to the coupled system of equations (Eq. 12) is

$$\begin{cases} d_{ux}^p = \sum_f S_x w_P, & o_{ux}^p = \sum_f S_x w_N \\ d_{uy}^p = \sum_f S_y w_P, & o_{uy}^p = \sum_f S_y w_N \\ d_{uz}^p = \sum_f S_z w_P, & o_{uz}^p = \sum_f S_z w_N \end{cases} \tag{15}$$

The velocity is coupled to the pressure field via the continuity equation (Eq. 1). The coupling can be derived from the Rhie-Chow interpolation technique formulated for co-located grids to avoid checkerboard fields [6]. According to this method, the velocity at cell faces can be interpolated as [6]

$$\mathbf{u}_f = \overline{\mathbf{u}} + \dfrac{1}{\overline{d_u^u}} \left( \overline{\nabla p} - \nabla p|_f \right) \tag{16}$$

where $d_u^u$ represent the set of diagonal coefficients ($d_{ux}^{ux}, d_{uy}^{uy}$ and $d_{uz}^{uz}$) of the momentum equation (Eq. 2). The terms with an overbar result from linear interpolation (from cell centers to face centers), whereas the last term in Eq. (16) is evaluated directly at cell faces.

When Eq. (16) is used to evaluate the discrete continuity equation (Eq. 1), the following relation is obtained [6],

$$\int_{V_P} \nabla \cdot \mathbf{u}\, dV_P = \sum_f \mathbf{S}_f \cdot \mathbf{u}_f = \sum_f \mathbf{S}_f \cdot \left[ \overline{\mathbf{u}} + \frac{1}{\overline{d_u^u}} \left( \overline{\nabla p} - \nabla p|_f \right) \right] = 0$$

$$\Leftrightarrow \sum_f \frac{\mathbf{S}_f \cdot \nabla p|_f}{\overline{d_u^u}} = \sum_f \mathbf{S}_f \cdot \overline{\mathbf{u}} + \sum_f \frac{\mathbf{S}_f \cdot \overline{\nabla p}}{\overline{d_u^u}}$$

(17)

Eq. (17) is the Poisson equation for pressure, where the left-hand-side is the semi-discretized Laplace operator (contributes to $a_p^p$ and $b_p$ in Eq. 12), the first term in the right-hand-side is the element coupling the velocity to pressure and the last term is added to the source vector (contributes to $b_p$ in Eq. 12). The coupling term is computed implicitly,

$$\sum_f \mathbf{S}_f \cdot \overline{\mathbf{u}} = \sum_f S_x \left( u_{x,P} w_P + u_{x,N} w_N \right) + S_y \left( u_{y,P} w_P + u_{y,N} w_N \right) + S_z \left( u_{z,P} w_P + u_{z,N} w_N \right) \quad (18)$$

and its contribution to the coupled system of equations (Eq. 12) is

$$\begin{cases} d_p^{ux} = -\sum_f S_x w_P, & o_p^{ux} = -\sum_f S_x w_N \\ d_p^{uy} = -\sum_f S_y w_P, & o_p^{uy} = -\sum_f S_y w_N \\ d_p^{uz} = -\sum_f S_z w_P, & o_p^{uz} = -\sum_f S_z w_N \end{cases} \quad (19)$$

### 3.1.2. Coupling between polymeric extra-stresses and velocity

The pressure-velocity coupling presented above has been discussed and used in several works related with Newtonian fluids (e.g. [6, 8, 13]), but the extra-stresses-velocity coupling is still an immature topic of discussion. To the best of our knowledge, this subject has been only recently addressed in [10], where the extra-stresses couple to velocity through the implicit discretization of term $\nabla \cdot \boldsymbol{\tau}$ in the momentum equation (Eq. 2),

$$\int_{V_P} \nabla \cdot \boldsymbol{\tau}\, dV_P = \sum_f \mathbf{S}_f \cdot \boldsymbol{\tau}_f$$

$$= \sum_f \begin{bmatrix} S_x(\tau_{xx,P} w_P + \tau_{xx,N} w_N) + S_y(\tau_{xy,P} w_P + \tau_{xy,N} w_N) + S_z(\tau_{xz,P} w_P + \tau_{xz,N} w_N) \\ S_x(\tau_{xy,P} w_P + \tau_{xy,N} w_N) + S_y(\tau_{yy,P} w_P + \tau_{yy,N} w_N) + S_z(\tau_{yz,P} w_P + \tau_{yz,N} w_N) \\ S_x(\tau_{xz,P} w_P + \tau_{xz,N} w_N) + S_y(\tau_{yz,P} w_P + \tau_{yz,N} w_N) + S_z(\tau_{zz,P} w_P + \tau_{zz,N} w_N) \end{bmatrix} \quad (20)$$

such that the contribution to the coupled system of equations (Eq. 12) is

$$\begin{cases} d_{ux}^{\tau xx} = d_{uy}^{\tau xy} = d_{uz}^{\tau xz} = -\sum_f S_x w_P, & o_{ux}^{\tau xx} = o_{uy}^{\tau xy} = o_{uz}^{\tau xz} = -\sum_f S_x w_N \\ d_{ux}^{\tau xy} = d_{uy}^{\tau yy} = d_{uz}^{\tau yz} = -\sum_f S_y w_P, & o_{ux}^{\tau xy} = o_{uy}^{\tau yy} = o_{uz}^{\tau yz} = -\sum_f S_y w_N \\ d_{ux}^{\tau xz} = d_{uy}^{\tau yz} = d_{uz}^{\tau zz} = -\sum_f S_z w_P, & o_{ux}^{\tau xz} = o_{uy}^{\tau yz} = o_{uz}^{\tau zz} = -\sum_f S_z w_N \\ d_{ux}^{\tau yy} = d_{ux}^{\tau yz} = d_{ux}^{\tau zz} = d_{uy}^{\tau xx} = d_{uy}^{\tau xz} = d_{uy}^{\tau zz} = d_{uz}^{\tau xx} = d_{uz}^{\tau xy} = d_{uz}^{\tau yy} = 0 \\ o_{ux}^{\tau yy} = o_{ux}^{\tau yz} = o_{ux}^{\tau zz} = o_{uy}^{\tau xx} = o_{uy}^{\tau xz} = o_{uy}^{\tau zz} = o_{uz}^{\tau xx} = o_{uz}^{\tau xy} = o_{uz}^{\tau yy} = 0 \end{cases} \quad (21)$$

In the same work [10], the velocity is coupled to extra-stresses by a *hybrid* discretization of the convective term in the constitutive equation. However, in the present work a different term of the constitutive equation (Eq. 3) is selected to couple velocity to extra-stresses, $\eta_P(\nabla \mathbf{u} + \nabla \mathbf{u}^T)$. This term, or a variation of it with a shear-rate-dependent scalar coefficient, is present in a number of viscoelastic models (Oldroyd-B, Giesekus, FENE-P, etc.), such that this coupling term can also be used in those cases. The implicit discretization of tensor $\eta_P \nabla \mathbf{u}$ results in the following matrix,

$$\int_{V_P} \eta_P \nabla \mathbf{u} \, dV_P = \eta_P \sum_f \mathbf{S}_f \otimes \mathbf{u}_f$$

$$= \eta_P \sum_f \begin{bmatrix} S_x(u_{x,P} w_P + u_{x,N} w_N) & S_x(u_{y,P} w_P + u_{y,N} w_N) & S_x(u_{z,P} w_P + u_{z,N} w_N) \\ S_y(u_{x,P} w_P + u_{x,N} w_N) & S_y(u_{y,P} w_P + u_{y,N} w_N) & S_y(u_{z,P} w_P + u_{z,N} w_N) \\ S_z(u_{x,P} w_P + u_{x,N} w_N) & S_z(u_{y,P} w_P + u_{y,N} w_N) & S_z(u_{z,P} w_P + u_{z,N} w_N) \end{bmatrix} \quad (22)$$

and the transpose of this expression is valid for the discrete $\eta_P \nabla \mathbf{u}^T$, which, upon addition, originate a symmetric tensor. The contribution of the complete term to the coupled system of equations (Eq. 12) is

$$\begin{cases} d_{\tau xx}^{ux} = -2\eta_P \sum_f S_x w_P, & o_{\tau xx}^{ux} = -2\eta_P \sum_f S_x w_N \\ d_{\tau yy}^{uy} = -2\eta_P \sum_f S_y w_P, & o_{\tau yy}^{uy} = -2\eta_P \sum_f S_y w_N \\ d_{\tau zz}^{uz} = -2\eta_P \sum_f S_z w_P, & o_{\tau zz}^{uz} = -2\eta_P \sum_f S_z w_N \\ d_{\tau xy}^{uy} = d_{\tau xz}^{uz} = -\eta_P \sum_f S_x w_P, & o_{\tau xy}^{uy} = o_{\tau xz}^{uz} = -\eta_P \sum_f S_x w_N \\ d_{\tau xy}^{ux} = d_{\tau yz}^{uz} = -\eta_P \sum_f S_y w_P, & o_{\tau xy}^{ux} = o_{\tau yz}^{uz} = -\eta_P \sum_f S_y w_N \\ d_{\tau xz}^{ux} = d_{\tau yz}^{uy} = -\eta_P \sum_f S_z w_P, & o_{\tau xz}^{ux} = o_{\tau yz}^{uy} = -\eta_P \sum_f S_z w_N \\ d_{\tau xx}^{uy} = d_{\tau xx}^{uz} = d_{\tau xy}^{uz} = d_{\tau xz}^{uy} = d_{\tau yy}^{ux} = d_{\tau yy}^{uz} = d_{\tau yz}^{ux} = d_{\tau zz}^{ux} = d_{\tau zz}^{uy} = 0 \\ o_{\tau xx}^{uy} = o_{\tau xx}^{uz} = o_{\tau xy}^{uz} = o_{\tau xz}^{uy} = o_{\tau yy}^{ux} = o_{\tau yy}^{uz} = o_{\tau yz}^{ux} = o_{\tau zz}^{ux} = o_{\tau zz}^{uy} = 0 \end{cases} \quad (23)$$

It is worth noting that the coupling between the extra-stress and velocity is naturally achieved by the implicit discretization of the terms present in the governing equations.

However, this procedure cannot be directly applied when some transformations of variable are applied to the constitutive equation such as, for example, in the log-conformation tensor approach [32]. Therefore, other coupling techniques are still needed in such cases.

### 3.1.3. Coupling between electric potential and species concentration

For the PNP system of equations, the coupling between species concentration and electric potential in the Poisson equation (Eqs. 5 and 7) is easily imposed by the implicit discretization of its right-hand-side, which would be explicitly computed in a segregated solution method. Thus, the contribution to the coupled system of equations (Eq. 12) is (assuming two ionic species)

$$\begin{cases} d_{\Psi}^{c+} = V_P F z_+, & o_{\Psi}^{c+} = 0 \\ d_{\Psi}^{c-} = V_P F z_-, & o_{\Psi}^{c-} = 0 \end{cases} \quad (24)$$

where $V_P$ is the volume of each cell **P**.

In the equations governing the transport of charged species, the electric potential is coupled to the species concentration through the electromigration term, which becomes a standard Laplace operator discretized implicitly in variable $\Psi$,

$$\int_{V_P} \nabla \cdot (\alpha c_i \nabla \Psi) dV_P = \alpha \sum_f c_{i,f} \mathbf{S}_f \cdot \nabla \Psi_f = \alpha \sum_f |\mathbf{S}_f| (c_{i,P} w_P + c_{i,N} w_N) \frac{\Psi_P - \Psi_N}{|\mathbf{x}_P - \mathbf{x}_N|} \quad (25)$$

with $\alpha = D_i \frac{e z_i}{kT}$. The final form taken by Eq. (25) assumes an orthogonal grid; an explicit correction term needs to be added for non-orthogonal grids [2], which is incorporated in the source vector (contributes to $b_{ci}$ in Eq. 12). The contribution to the matrix of coefficients in the coupled system of equations (Eq. 12) is (assuming two ionic species)

$$\begin{cases} d_{c+}^{\Psi} = \dfrac{-\alpha \sum_f |\mathbf{S}_f|(c_{+,P} w_P + c_{+,N} w_N)}{|\mathbf{x}_P - \mathbf{x}_N|}, & o_{c+}^{\Psi} = \dfrac{\alpha \sum_f |\mathbf{S}_f|(c_{+,P} w_P + c_{+,N} w_N)}{|\mathbf{x}_P - \mathbf{x}_N|} \\ d_{c-}^{\Psi} = \dfrac{-\alpha \sum_f |\mathbf{S}_f|(c_{-,P} w_P + c_{-,N} w_N)}{|\mathbf{x}_P - \mathbf{x}_N|}, & o_{c-}^{\Psi} = \dfrac{\alpha \sum_f |\mathbf{S}_f|(c_{-,P} w_P + c_{-,N} w_N)}{|\mathbf{x}_P - \mathbf{x}_N|} \end{cases} \quad (26)$$

Note that the coupling between electric potential and species concentration does not apply for the PB model. In this model, the two Poisson equations for the electric potential variables (Eqs. 9 and 10) do not depend on any other variable other than the electric potential itself and are solved individually, in sequence.

### 3.1.4. Coupling between electric potential and velocity

The coupling in EDF can be strengthened even more with the implicit discretization of the electric body force in the momentum equation (Eq. 2), thus coupling the electric potential to the velocity. The volume integration of this body force can then be written as

$$-\int_{V_P} \rho_E \nabla \Psi \, dV_P = -\rho_{E,P} \sum_f \mathbf{S}_f \Psi_f = -\rho_{E,P} \sum_f \begin{bmatrix} S_x \\ S_y \\ S_z \end{bmatrix} (\Psi_P w_P + \Psi_N w_N) \tag{27}$$

which is similar to the Green-Gauss theorem applied in Eq. (13) to the pressure gradient, except for a multiplicative cell-dependent term ($\rho_{E,P}$). Note that this (explicit) multiplicative term is accounted for in such a way to recover exactly the result that would be obtained for an explicit discretization of the body-force term (as in a segregated solver), and differs, for example, from the linear interpolation used for the multiplicative $c_i$ term in Eq. (25). The contribution of this term to the coupled system of equations (Eq. 12) is

$$\begin{cases} d_{ux}^{\Psi} = \rho_{E,P} \sum_f S_x w_P, & o_{ux}^{\Psi} = \rho_{E,P} \sum_f S_x w_N \\ d_{uy}^{\Psi} = \rho_{E,P} \sum_f S_y w_P, & o_{uy}^{\Psi} = \rho_{E,P} \sum_f S_y w_N \\ d_{uz}^{\Psi} = \rho_{E,P} \sum_f S_z w_P, & o_{uz}^{\Psi} = \rho_{E,P} \sum_f S_z w_N \end{cases} \tag{28}$$

The coupling between velocity and electric-related variables envisaged in Eq. (27) is not unique, as we could have used implicit $c_i$ (incorporated in $\rho_E$) and explicit $\Psi$ to impose the coupling, or a combination of the two methods. Eq. (27) can be applied to both PNP and PB models. However, in this work we only apply it to the PNP model, since the electric potential in the PB model quickly stabilizes after few iterations (one iteration for orthogonal grids) from the beginning of the simulation and the electric body-force simply becomes a steady source term in the momentum equation.

### 3.2. Semi-coupled solver

Depending on the particular flow problem, the implicit coupling (at the matrix level) between some pairs of variables contributes more for stability and time accuracy than other pairs. Given that each coupling relation increases the size and complexity of the final coupled system of equations (Eq. 12), it is sometimes beneficial to drop some of these coupling terms. In this work, the solvers that do not use all the implicit coupling terms described in the previous sections, and represented in Eq. (12), are named semi-coupled solvers. Among the several combinations that could result, we will only explore two of them in the current study. One of the semi-coupled solvers splits the Navier-Stokes plus constitutive equations from the PNP equations. Thus, $p$-**u**-**τ** are solved coupled, but separated from $\Psi$-$c_i$, which are solved coupled in their own system of equations. In the other semi-coupled solver, **τ** is further split from the $p$-**u**-**τ** system of equations, such that each $\tau_{ij}$ component is solved individually and decoupled from the velocity and pressure.

### 3.3. Segregated solver

The segregated solution method has been presented and extensively used in [15, 27]. We will not elaborate on this method, but we shall note that the SIMPLEC algorithm [33] is used for pressure-velocity coupling. This semi-implicit method does not require under-relaxation of the pressure variable [33]. The explicitness of the segregated solution method can be decreased by inner-iterating all the equations multiple times within the same time-step [15, 27]. In this process, the equations are solved repeatedly inside a loop and the explicit terms are updated with the solution from the previous iteration. This increases both stability and time accuracy.

### 3.4. Discretization schemes for non-coupling terms

For the terms whose discretization was not previously described, the spatial derivatives are discretized using centered differences and the convective terms are discretized with the high-resolution CUBISTA scheme [34], implemented according to a component-wise, deferred correction approach [27]. The algorithm is second-order accurate in space, as demonstrated in [15, 27] for segregated solvers. The coupling terms are discretized with the same schemes in both coupled and segregated solvers, as previously mentioned. Therefore, the spatial order of accuracy remains unchanged for the coupled solvers.

For transient simulations, time-derivatives are discretized with the three-time level scheme [27]. The modifications introduced by coupled solvers do not affect the discretization of time-derivatives, which, therefore, retain their formal second-order accuracy. However, since much of the explicitness of segregated solvers is removed by the coupled solution method, the overall error in the temporal dimension is generally smaller for coupled solvers, considering similar test conditions. Nevertheless, the PNP system of equations still needs to be self-iterated at least twice to retain second-order accuracy in time [14, 15], which was assessed in the benchmark proposed in [15] (results not shown for conciseness). This is due to the explicit $c_i$ used in the coupling term of Eq. (25); only a fully-implicit electromigration term would allow second-order accuracy in a single iteration scenario.

For steady-state simulations, time accuracy is not an issue of concern. The obvious procedure to simulate steady-state flows is removing time-derivatives from the governing equations. If all the remaining terms were implicitly discretized and if a coupled solution method was applied, convergence would be reached in a single iteration (time-step). However, the existence of explicit terms deteriorates the rate of convergence, even more in segregated solvers. Moreover, the removal of time-derivatives from the equations leads frequently to numerical divergence. This is either due to insufficient diagonal dominance in the matrices being solved, or due to the fast change of the solution, combined with the explicitness of some terms. Thus, in practice, there is a need for a controlled (slower) evolution to the steady-state solution. This can be achieved by retaining the time-derivatives and using relatively large time-steps. The implicit backward Euler scheme is a good option to use for the discretization of the time-derivatives in such cases, where stability is preferred over accuracy. Another popular method for steady-state simulations of non-linear problems consists in the use of under-relaxation [2]. In this procedure, a variable is only allowed to change by a fraction of its variation without under-relaxation, which is controlled by under-relaxation factors ($\alpha$). In practice, under-relaxation increases the diagonal dominance of a matrix by scaling its diagonal coefficients. Time-stepping and under-relaxation work in a similar manner for steady-state calculations and the choice between them is a compromise between numerical stability and convergence rate, and depends not only on the equations to which they are applied, but also on the specific problem being solved. For example, during this work we observed that time-stepping tends to be more efficient for the species transport equation and constitutive equation, whereas under-relaxation seems to be a better option for the momentum equation. In some cases, the combination of both methods is the most efficient option.

### 3.5. Solution of linear systems of equations

The coupled system of equations (Eq. 12) solved in this work can have up to 13 more rows and columns than each individual system of equations assembled and solved in the segregated solution method. If the matrix solving stage already is the bottleneck of most implicit CFD codes implementing segregated solvers, then this is expected to be worse for coupled solvers. The high memory usage and high computational cost *per* time-step of coupled solvers in the matrix solving stage has been the principal obstacle to its widespread use in FVM.

Consider the linear system of equations $\mathbf{Ax} = \mathbf{b}$, where $\mathbf{A}$ represents the matrix of coefficients, $\mathbf{b}$ is the right-hand-side vector containing the source terms of the discretized equations, and $\mathbf{x}$ is the solution vector. There are essentially two main classes of methods that can be used to compute $\mathbf{x}$: direct and iterative solvers.

Direct solvers can find the *exact* solution $\mathbf{x}$, i.e. $|\mathbf{Ax} - \mathbf{b}|$ evaluates to machine precision in a finite number of operations. For example, a direct solver based on the LU decomposition first factorizes matrix $\mathbf{A}$ into lower/upper triangular matrices $\mathbf{L}$ and $\mathbf{U}$, such that $\mathbf{LU} = \mathbf{A}$, and then solves two triangular systems of equations: $\mathbf{Ly} = \mathbf{b}$ and $\mathbf{Ux} = \mathbf{y}$ using, for example, backward/forward substitution. If the coefficients of matrix $\mathbf{A}$ do not change over time, the factorization needs only to be performed once and both triangular matrices $\mathbf{L}$ and $\mathbf{U}$ can be reused for different $\mathbf{b}$ vectors. This option saves time at the cost of increasing memory usage. Direct solvers are very robust, but present a high memory usage, even when specifically adapted for sparse matrices. They are typically used for grids of small to medium size, being very popular among finite-element methods [35].

Iterative solvers start from an initial guess of $\mathbf{x}$, which is then iteratively improved by minimizing a residual (for example, $|\mathbf{Ax} - \mathbf{b}|$). The tolerance defined for the residuals determines the accuracy of the final solution. Multigrid and Krylov subspace solvers are among the most popular types of iterative solvers. These methods suffer generically from convergence problems, for which they are typically used along with a preconditioning technique. The type and quality of the preconditioner used is often more important than the iterative solver itself regarding the convergence rate. Considering left preconditioning, the original system of equations is transformed into $\mathbf{M}^{-1}\mathbf{Ax} = \mathbf{M}^{-1}\mathbf{b}$ after multiplication by matrix $\mathbf{M}^{-1}$ from which the preconditioner is derived [36]. The resulting preconditioned system should have a lower condition number and, therefore, the iterative solver converges faster. The closer $\mathbf{M}$ is to $\mathbf{A}$, the faster will be the convergence, but more time will be spent in the computation of $\mathbf{M}^{-1}$. In the same way direct solvers can reuse the matrix factorization if the matrix of coefficients does not change over time, iterative solvers can also reuse the preconditioner. This option is most advantageous when $\mathbf{M}$ is a good preconditioner, but costly to compute, although it incurs in a memory overhead. Comparing to direct solvers, iterative solvers are much more sensitive to the condition number of $\mathbf{A}$, which is problem-dependent, but use fewer memory resources. Therefore, they are particularly suitable for problems with meshes having a large number of cells, being a common choice in FVM. For example, there is a number of iterative solvers available in OpenFOAM® for sparse matrices, but no direct solver is currently available for such matrices.

Previous works applying coupled solvers in the FVM used mostly multigrid [6, 8, 9, 13] or preconditioned iterative solvers [11]. In this work, we resort mainly on direct solvers (LU factorization), when the matrix of coefficients does not change after the first iteration, and an iterative solver (BiCGStab [37]) combined with an LU preconditioner otherwise. The use of an iterative solver preconditioned with a complete LU factorization might seem counterintuitive. This setup is nearly equivalent to a direct solver and convergence is achieved in a single iteration of the iterative solver, as long as the LU factors result from the matrix **A** being solved. However, in opposition to a direct solver, this approach allows to apply the LU factors to a matrix **A** different from the one from which they were computed, i.e. it is possible to reuse the preconditioner. The main criteria used to select those solvers was robustness, but also ease of use (few adjustable parameters, comparing for example with multigrid methods). However, to make these solvers efficient, the factorization/preconditioner had to be reused during the simulations. For example, the matrix of coefficients for the $p$-**u** coupled system of equations does not change over time for constant viscosity, constant time-step (or constant under-relaxation factor) and creeping-flow conditions (zero Reynolds number, which is achieved by dropping out the convective term in the momentum equation). Hence, the factorization/preconditioner only needs to be computed once. On the other hand, for the coupled systems $p$-**u**-**τ**, $\Psi$-$c_i$ and $p$-**u**-**τ**-$\Psi$-$c_i$, the matrix of coefficients changes every iteration due to the convective and/or coupling terms, but the same preconditioner can sometimes be used for more than one iteration/time-step. This is the case, for example, when the matrix of coefficients have a slow variation over time and/or when they stabilize after a few iterations/time-steps. In these situations, a simple empirical expression was devised to automatically decide when the preconditioner should be updated. If we consider $n$ as the number of time-steps (after the current one) in which the actual preconditioner can be reused, then

$$n = \min\left\{10^5, \text{int}\left[\frac{t_{\text{CPU},0}}{t_{\text{CPU}}} \frac{1}{\left|1 - \frac{t_{\text{CPU},1}}{t_{\text{CPU}}} + 10^{-6}\right|}\right]\right\} \qquad (29)$$

In Eq. (29), $t_{CPU}$ is the CPU time required to solve the matrix at a given time-step, $t_{CPU,0}$ is the CPU time required to solve the matrix when the preconditioner is computed (includes the time spent to compute the preconditioner) and $t_{CPU,1}$ is the CPU time required to solve the matrix, reusing the last pre-conditioner computed, in the first time-step after which the preconditioner has been updated. Note that $t_{CPU}$ is computed every time-step, whereas $t_{CPU,0}$ and $t_{CPU,1}$ are only updated each time the preconditioner is updated. Unless otherwise stated, when the preconditioner is reused but needs to be updated periodically (matrix of coefficients changes over time), Eq. (29) is used for such purpose (the preconditioner is always computed in the first three iterations/time-steps of a given simulation, thus initializing Eq. 29).

In the FVM literature, several definitions can be found for the residual of a system of equations, or, more simply, for the residual associated to the variable being solved. In this work, we use the definition implemented by default in OpenFOAM®,

$$\text{Residual} = \frac{|\mathbf{Ax} - \mathbf{b}|_1}{|\mathbf{Ax} - \mathbf{A}\bar{x}\mathbf{1}|_1 + |\mathbf{b} - \mathbf{A}\bar{x}\mathbf{1}|_1} \tag{30}$$

where $\bar{x}$ denotes the average value of the solution vector, $\mathbf{1}$ is a vector of ones and $| \ |_1$ represents the $L_1$-norm. This definition applies directly to segregated solvers and is applied individually to each equation inserted in a coupled system of equations, such that the residual for each component of a field variable can be extracted. Unless otherwise stated, a steady-state problem is considered converged in this work when the residuals of all variables being solved drop below $10^{-6}$.

The absolute and relative tolerances of iterative solvers are set to $10^{-9}$ and 0, respectively (in practice, only the former is used). For direct solvers, the systems of equations are solved to machine precision. Note that such software-specific tolerances are related with the error (residual) associated to iterative linear solvers (matrix solving), and shall not be confused with Eq. (30) (field residuals).

### 3.6. Linking OpenFOAM® to PETSc

The default sparse matrix solvers available in OpenFOAM® are specifically prepared to handle the OpenFOAM® *ldu* matrix format, which is intrinsically linked to the mesh structure. This setup is naturally efficient but presents low flexibility. For example, it does not allow reusing preconditioners, and an *ldu* matrix can only handle a single variable. Moreover, direct solvers for sparse matrices are not implemented and the preconditioners currently available are not sufficiently robust in some situations, leading to slow convergence of the iterative solvers. Thus, given these limitations, we implemented interfaces to link OpenFOAM® to external libraries offering efficient sparse matrix solvers: Eigen [38], Hypre [39] and PETSc [40-42]. The three libraries are open-source, in active development and oriented to achieve high-performance, they have a long track record and they can be easily integrated in the OpenFOAM® framework. The Eigen library is probably the simplest one to use, but it is also the only not making use of the MPI parallelism implemented in OpenFOAM® (some solvers can use OpenMP). Hypre and PETSc are both directed to achieve high performance in massively parallel systems, but PETSc offers a wider variety of sparse matrix solvers/preconditioners and related utilities. Thus, the machinery required to assemble and solve the coupled system of equations in multiple processors has been implemented exclusively for the PETSc interface, although the three interfaces can be used to solve any generic single-variable sparse matrix assembled in OpenFOAM®. For the sake of conciseness, only the PETSc interface will be briefly described, but all the three interfaces are available in rheoTool [20].

PETSc is a library implementing a set of scalable routines that can be used to solve partial differential equations [41]. A number of matrix formats, solvers (direct and iterative) and preconditioners are available, as well as interfaces to other packages (it is even possible to use Hypre from PETSc) [41]. Parallelism is supported via MPI (the only explored in this work) and OpenMP on CPUs, and it is also possible to use GPU accelerators [41].

The sequence of operations connecting OpenFOAM® and PETSc is schematized in Fig. 1. For an arbitrary field *p*, the equation implicit in this variable is assembled in OpenFOAM® in a *fvMatrix* container. The *fvMatrix* holds all the information needed to build matrix **A** and the source vector **b** in PETSc formats. This operation requires an element-wise transfer of data between libraries. The linear system of equations is then solved by PETSc using any of the runtime selectable methods available and the solution vector (**x**), in PETSc format, is copied element-wise to the field container in OpenFOAM® format. The final step is the cleaning of all temporary structures that have been created in PETSc (**A**, **b**, **x** and the linear solver containing, among others, the preconditioner). This step can be optionally skipped at intermediary time-steps if the structures are intended to be reused in the next time the equation for the same variable is solved for.

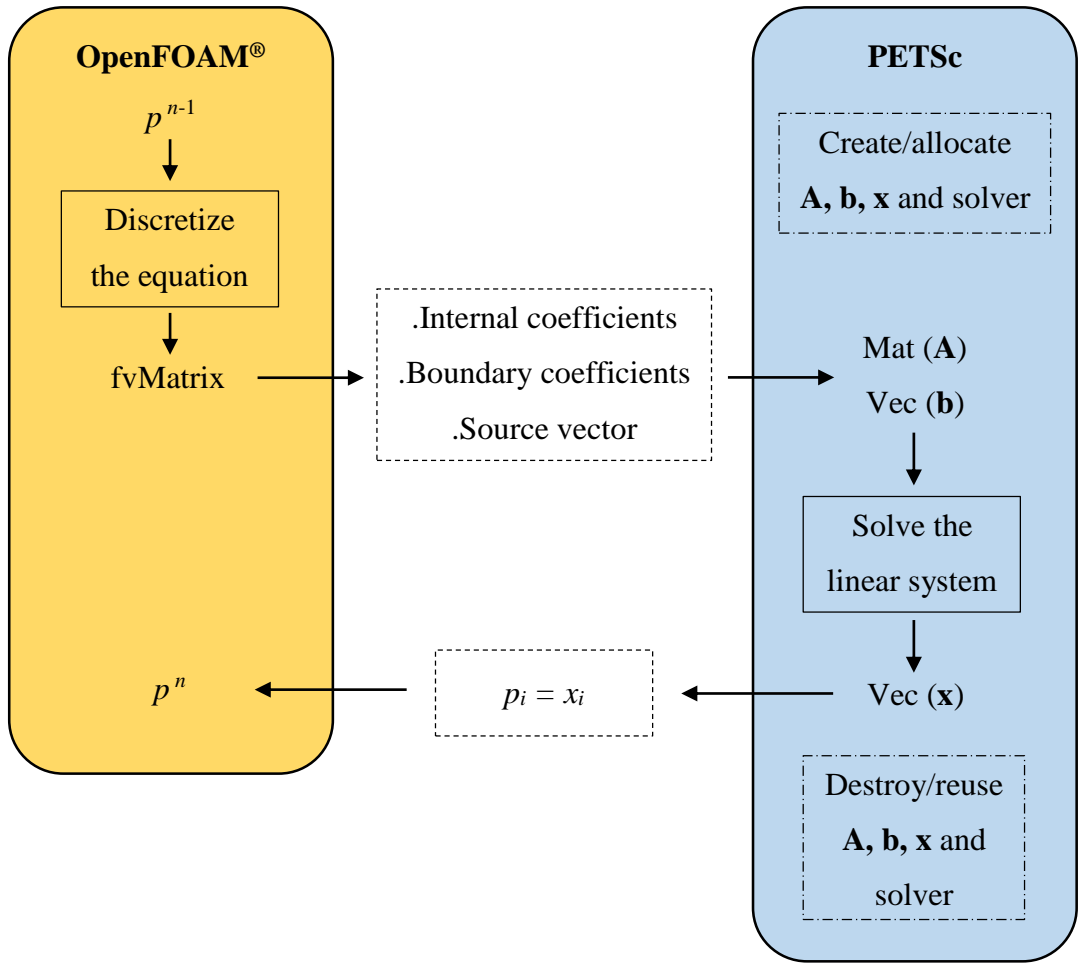

**Figure 1** – Matrix solving stage using the OpenFOAM®–PETSc interface. The equation for an arbitrary field $p$ is assembled in OpenFOAM® and its coefficients are transferred to PETSc, which assembles and solves the same system and transfers the solution back to OpenFOAM®.

## 4. Hardware and software specifications

The simulations performed in this work were carried out in a workstation (Hewlett-Packard 158B motherboard) equipped with 2 Intel® Xeon® E5-2643 processors (3.30 GHz , 8 cores, 16 threads) and 64 GB of available RAM memory, running under a Ubuntu 18.04.01 operating system (64 bits). OpenFOAM® version 6.0 is used in combination with PETSc release 3.10.2, both running with double precision. The MPI communication is ensured with OpenMPI 2.2.1 and OpenMP is disabled.

The direct LU solver used in this work via PETSc is from MUMPS package [43, 44]. In addition, the default parameters are retained in all sparse matrix solvers used in this work, whether they are from OpenFOAM® or PETSc (only the absolute and relative tolerances are adjusted).

## 5. Results

In this section, we start by verifying the coupled solver against an analytical solution for the electroosmotic flow of a PTT fluid in a microchannel. Then, the performance of the coupled solvers is assessed in two other EDF. Unless otherwise stated, all the simulations were carried out with a single processor, using the sparse matrix solvers from PETSc library. The following abbreviations are used in this section referring to sparse matrix solvers and preconditioners: GAMG (Geometric agglomerated Algebraic MultiGrid), CG (Conjugate Gradient), BiCG (Bi-Conjugate Gradient), BiCGStab (Bi-Conjugate Gradient Stabilized), DIC (Diagonal Incomplete Cholesky) and DILU (Diagonal Incomplete LU).

### 5.1. Solver verification: electroosmotic flow of a PTT fluid in a microchannel

The main purpose of this section is to compare the results of the coupled solver with an analytical solution, in order to confirm that the implicit discretization of the coupling terms was correctly implemented. The EDF selected for such purpose is the electroosmotic flow of a linear PTT fluid in a straight, two-dimensional microchannel [45].

The computational domain is a segment of the cross-section of a 2D microchannel ($xy$ plane), where $H$ is the half-width of that cross-section. Since only the fully-developed flow region is of interest, the mesh has only one cell in the streamwise direction ($x$-direction) and 600 cells are used in the cross-stream direction (the minimum cell size normal to the walls is $\lambda_D/30$).

Considering the PNP model for ions transport, periodic boundary conditions were assigned in the streamwise direction, whereas at the wall no-slip and no-penetration, zero flux of ionic species, fixed intrinsic potential ($\psi_w$) and linear extrapolation of extra-stress components [27] were imposed. The flow is driven by a uniform electric field of magnitude $E$, applied in the direction tangential to the walls, and also by a pressure-gradient ($p'$) acting in the same direction, but not necessarily in the same sense as the electric field.

The analytical solution presented in [45] depends on the ratio between the pressure gradient and the applied electric field, $\Gamma = -\dfrac{H^2 p'}{\psi_w \varepsilon_0 \varepsilon_r E}$, the relative size of the EDL, $\tilde{\kappa} = \dfrac{H}{\lambda_D}$, and the modified Deborah number, $\sqrt{\varepsilon}De_\kappa = \sqrt{\varepsilon}\dfrac{\lambda U}{\lambda_D}$, where the velocity scale is given by $U = -\dfrac{\varepsilon_0 \varepsilon_r \psi_w E}{\eta_P}$. Moreover, the analytical solution is only valid for $\eta_S = 0$.

The numerical solution obtained with the coupled solver is plotted in Fig. 2 for $\tilde{\kappa} = 25$, $\sqrt{\varepsilon}De_\kappa = 5$ and $\Gamma \in [-4,4]$. As can be seen, a good agreement is observed between the numerical results and the analytical solution. Similar results were obtained with the segregated solver and/or by replacing the PNP model by the PB model (results not shown).

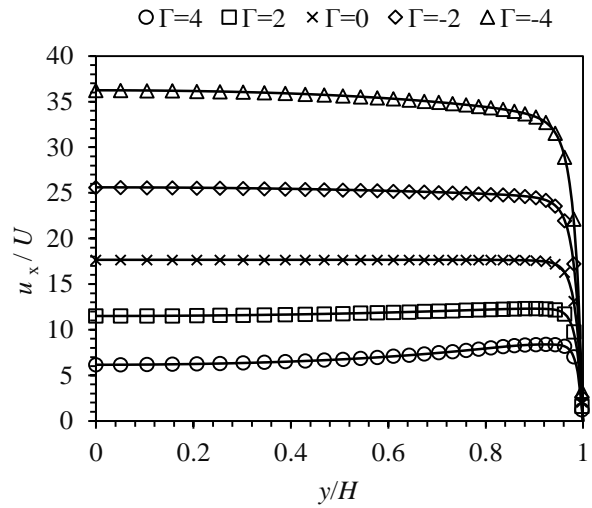

**Figure 2** – Velocity profiles for the electroosmotic flow of a simplified linear PTT fluid in a 2D microchannel, for $\tilde{\kappa} = 25$ and $\sqrt{\varepsilon}De_\kappa = 5$. The lines represent the analytical solution from [45] and the symbols correspond to the numerical solution obtained with the coupled solver developed in this work.

### 5.2. Case I: induced-charge electroosmosis around a metallic cylinder

The first test case selected is the ICEO of a Newtonian fluid around a metallic cylinder, using the PNP model for the transport of ionic species. At low voltages induced in the cylinder, the flow reaches a steady-state characterized by four counter-rotating vortices around the cylinder [46], whereas at high voltages this organized flow pattern breaks down and degenerates in a chaotic flow [19]. Besides being an important problem from the experimental/engineering point of view, the ICEO also allows to test and benchmark the coupled solver under different flow regimes. Moreover, as mentioned in the Introduction, the failure of the segregated solver to simulate chaotic ICEO has been the main motivation for this work.

The 2D computational domain is shown in Fig. 3 and consists of an infinitely long metallic cylinder (radius *b*) immersed in a theoretically unbounded fluid domain. In practice, the domain is bounded, but the boundaries are placed far enough (50$b$ from the cylinder center) to ensure minimal influence in the results. Two grids are used to discretize the computational domain, whose characteristics are presented in Table 1. The cells are compressed in the radial direction towards the cylinder surface, whereas they are uniformly distributed in the azimuthal direction.

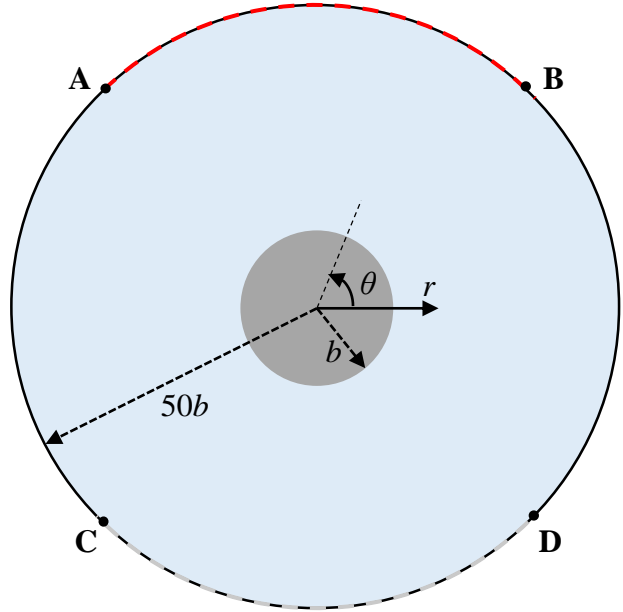

**Figure 3** – Schematic representation of the computational domain for the ICEO test case. The infinitely long metallic cylinder of radius $b$ is immersed in a domain of radius 50$b$ and a potential difference is applied between arcs **AB** and **CD**. The drawing is not to scale.

**Table 1** – Details of the two computational meshes used in the ICEO test case. The cells are compressed towards the cylinder surface, in the radial direction, whereas they are uniformly distributed in the azimuthal direction.

|  | Mesh | |
| --- | --- | --- |
|  | **M0** | **M1** |
| Number of cells in radial direction | 115 | 230 |
| Number of cells in azimuthal direction | 220 | 440 |
| Total number of cells | 25300 | 101200 |
| Minimum cell size in radial direction | $\lambda_D/50$ | $\lambda_D/100$ |
| Minimum cell size in azimuthal direction | $\pi b/110$ | $\pi b/220$ |

The following set of boundary conditions is used:

- arc(**AB**): $\Psi = V$, $\nabla p \cdot \mathbf{n} = 0$, $\mathbf{u} = \mathbf{0}$, $c_i = c_0$;
- arc(**CD**): $\Psi = -V$, $\nabla p \cdot \mathbf{n} = 0$, $\mathbf{u} = \mathbf{0}$, $c_i = c_0$;
- arc(**AC**) and arc(**BD**): $\nabla \Psi \cdot \mathbf{n} = 0$, $\nabla p \cdot \mathbf{n} = 0$, $\mathbf{u} = \mathbf{0}$, $c_i = c_0$;
- Cylinder surface: $\Psi = 0$, $\nabla p \cdot \mathbf{n} = 0$, $\mathbf{u} = \mathbf{0}$, $J_i = 0$.

where $J_i = \left( c_i \mathbf{u}_i - D_i \nabla c_i - c_i D_i \dfrac{e z_i}{kT} \nabla \Psi \right)_A \cdot \mathbf{S}_A$ represents the flux of species $i$ through surface A (cylinder surface), characterized by its surface-normal vector $\mathbf{S}_A = \mathbf{n}_A A_A$.

The simulations were performed considering two ionic species and $b = 10$ μm, $\eta = 0.001$ Pa.s, $\rho = 1000$ kg/m³, $\varepsilon_r = 84$, $T = 300$ K, $z_+ = -z_- = z = 1$, $D_+ = D_- = D = 10^{-9}$ m²/s and variable $c_0$ and $V$. The bulk concentration $c_0$ is adjusted in order to fix the Debye length or, equivalently, the relative size of the EDL, $\tilde{\kappa} = \dfrac{b}{\lambda_D}$. The applied voltage is set to impose a given value of $\tilde{\zeta} = \dfrac{Ebez}{kT}$, which represents the ratio between the induced potential on the cylinder surface and the thermal potential, where $E = \dfrac{V}{50b}$ is the electric field magnitude in the vicinity of the cylinder. Considering the velocity scale $U = \dfrac{\varepsilon_0 \varepsilon_r E^2 b}{\eta}$, the Reynolds number is defined as $Re = \dfrac{\rho U b}{\eta}$ and because it is lower than unity for all the range of parameters tested, the convective term is removed from the momentum equation.

The velocity values presented later in this section are normalized by the theoretical solution expected at steady-state for $\tilde{\zeta} \ll 1$ and $\tilde{\kappa} \gg 1$ [46],

$$\left( u_{r,\text{theo}}, u_{\theta,\text{theo}} \right) = -2U \left( \dfrac{b(b^2 - r^2)}{r^3} \cos 2\theta, \dfrac{b^3}{r^3} \sin 2\theta \right) \tag{31}$$

and time is scaled by a diffusive time-scale, $t_D = \dfrac{b^2}{D}$.

The semi-coupled solver used in this test case solves two separate coupled systems of equations: $p$-$\mathbf{u}$ and $\Psi$-$c_i$.

### 5.2.1. Steady-state solution at low voltages

We start the analysis by investigating the performance of the solvers to compute the steady-state solution at low voltage ($\tilde{\zeta} = 0.1$) and for two relative EDL sizes ($\tilde{\kappa} = 10$ and 100). The coupled and semi-coupled solvers required a very small amount of under-relaxation ($\alpha_u = 0.9999$) in the momentum equation to converge and no under-relaxation was needed in the species transport equation. On the other hand, the segregated solver failed to converge with such parameters. To avoid numerical divergence, we used $\alpha_u = 0.95$ and $\Delta t_{ci} = t_D/10^2$ ($\tilde{\kappa} = 10$) or $\Delta t_{ci} = t_D/10^4$ ($\tilde{\kappa} = 100$), i.e. the momentum equation was evolved with under-relaxation and the species transport equations were evolved through time-stepping (using a large, inaccurate time-step for each $\tilde{\kappa}$).

The residuals evolution is illustrated in Fig. 4 for one set of conditions (mesh M0, $\tilde{\kappa} = 10$ and $\tilde{\zeta} = 0.1$) and the number of iterations and computational time to convergence are listed in Table 2 for all the combinations of parameters tested. The criteria used to assess the convergence of a simulation was the drop of residuals below $10^{-6}$, which was found to be sufficient to ensure steady fields. The number of iterations to convergence is remarkably smaller for coupled and semi-coupled solvers, which converged in a dozen of iterations, whereas the segregated solver required more than one thousand iterations in most cases to achieve the same reduction of residuals. However, since the cost of each iteration is significantly higher for the coupled and semi-coupled solvers, in practice the gain of time (simulation speedup) is more modest and ranges from 3 to 17 (Table 2).

**Table 2** – Computational time and total number of iterations until convergence for the different solution methods tested in the simulation of the steady-state ICEO case ($\tilde{\zeta} = 0.1$). The values inside parentheses in the CPU time column represent the speedup factor relative to the segregated solution method using sparse matrix solvers available in OpenFOAM® (two last rows).

| Solution method | Matrix solvers | $\tilde{\kappa}$ | Total iterations | | Total CPU time (s) | |
|---|---|---|---|---|---|---|
| | | | Mesh M0 | Mesh M1 | Mesh M0 | Mesh M1 |
| Coupled | . $p$-**u**-$\Psi$-$c_i$: BiCGStab+LU (reuse) | 10 | 9 | 14 | 22 (3) | 144 (6) |
| | | 100 | 11 | 12 | 25 (4) | 150 (4) |
| Semi-coupled | . $p$-**u**: LU (reuse) . $\Psi$-$c_i$: BiCGStab+LU (reuse) | 10 | 6 | 10 | 9 (6) | 49 (17) |
| | | 100 | 6 | 7 | 9 (10) | 47 (11) |
| Segregated | . $p$, **u**: CG + DIC[(*)] . $\Psi$, $c_i$: GAMG[(*)] | 10 | 501 | 1646 | 56 | 848 |
| | | 100 | 1120 | 1443 | 94 | 535 |

(*) Sparse matrix solver from OpenFOAM®.

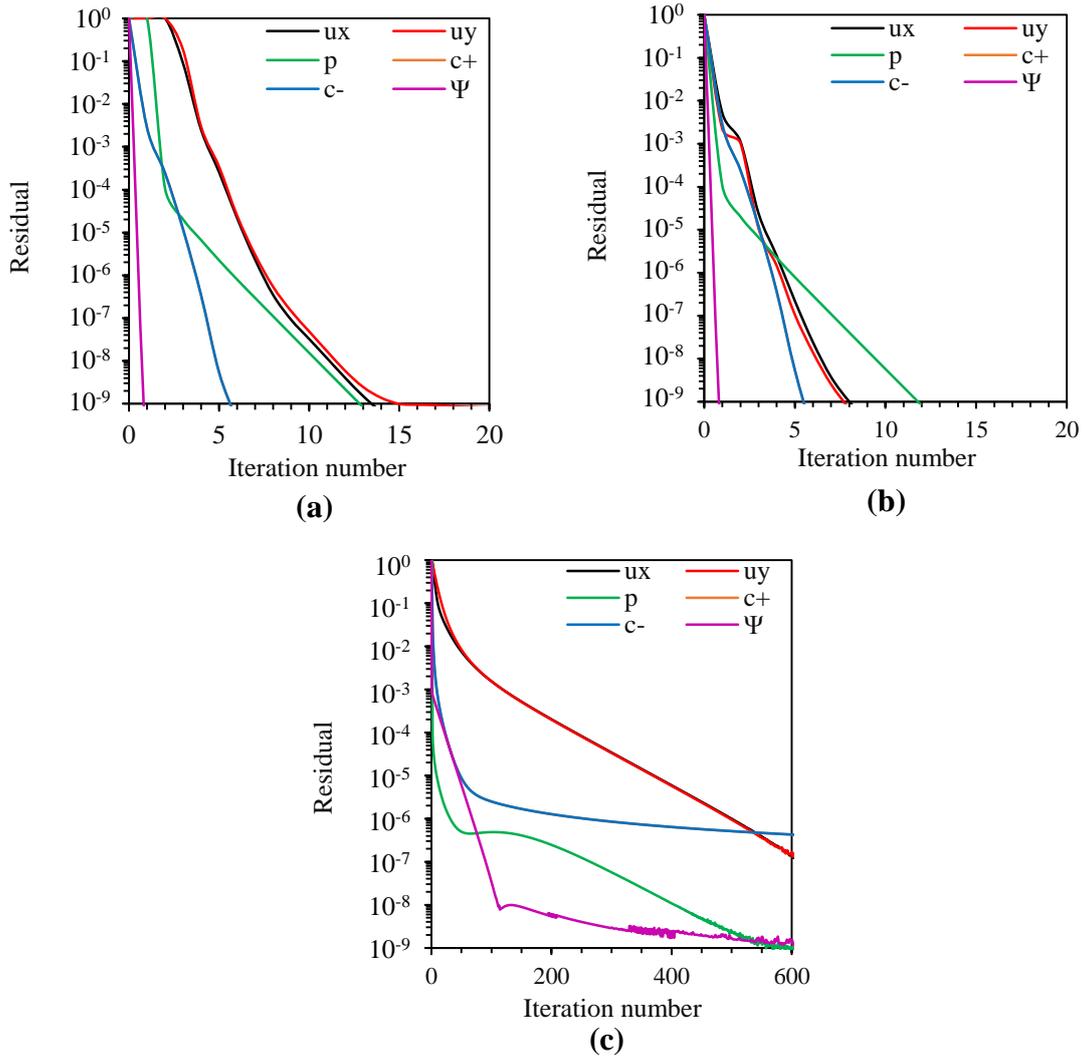

**Figure 4** – Residuals evolution for the steady-state ICEO case simulated in mesh M0 for $\tilde{\kappa} = 10$ and $\tilde{\zeta} = 0.1$, using (a) coupled, (b) semi-coupled and (c) segregated solvers. Note that the curves for $c_+$ and $c_-$ are virtually indistinguishable between them at the plotting scale.

Comparing between the coupled and semi-coupled solvers, we can observe that the total number of iterations is slightly smaller for the semi-coupled solver (Table 2). However, the main reason explaining the different speedup factors obtained is the computational time per time-step in each case, which is roughly half for the semi-coupled solver. This is not only due to differences in the complexity (size) of the matrices being solved in each case, but also due to different policies in the reuse of preconditioners/factorizations. For the coupled solver, the matrix of coefficients changes each time-step, such that the preconditioner can be only used for a limited number of time-steps (Eq. 29). In the semi-coupled solver, the coefficients of the coupled $p$-**u** equation do not change over time and the factorization of the direct LU solver computed in the first time-step can be reused during all the simulation under creeping flow conditions. The preconditioner for the coupled PNP system of equations still needs to be updated during the simulation (Eq. 29), since the matrix of coefficients is changing every time-step.

It is interesting to note in Fig. 4 that the residuals for $\Psi$ drop below the convergence threshold and stabilize from the first time-step for the coupled and semi-coupled solvers. This is because the equation for $\Psi$ is fully implicit in such solvers, and has no source terms (for an orthogonal grid). In addition, it can be seen in Table 2, for coupled and semi-coupled solvers, that the number of iterations to convergence is almost independent of the grid size for $\tilde{\kappa} = 100$, but shows some dependency for $\tilde{\kappa} = 10$. Additional tests at different $\tilde{\kappa}$ values revealed that low $\tilde{\kappa}$ values slightly impair the convergence rate as the mesh resolution is increased. However, coupled and semi-coupled solvers still clearly outperform the segregated solver regarding the grid size independency of the convergence rate.

### 5.2.2. Transient solution at mild voltages

After evaluating the performance of the coupled solver to reach the steady-state solution, now we evaluate the efficiency and accuracy to capture the transient behavior of a solution converging to steady-state. The benchmark variable used was the maximum azimuthal velocity ($u_{\theta,\max}$) along the radial direction at $\theta = 3\pi/4$. The voltage and ion concentration are fixed, such that $\tilde{\zeta} = 1$ and $\tilde{\kappa} = 10$. Unless otherwise stated, a single inner-iteration is performed in all cases, i.e. the governing equations are only solved once each time-step.

The time evolution of $u_{\theta,\max}$ is plotted in Fig. 5 for the different solution methods, tested with different time-steps. For the same time-step, it is clear that both the coupled and semi-coupled solvers are significantly more accurate than the segregated solver. Indeed, both the coupled and semi-coupled solvers can still approach the reference solution with a negligible error with $\Delta t = t_\mathrm{D}/100$, whereas a time-step 40 times smaller than this value is required for the segregated solver to reach a similar level of accuracy. The accuracy of the segregated solver can be also increased by increasing the number of inner iterations (curve "M0, *$\Delta$t = $t_\mathrm{D}$/200" in Fig. 5c was obtained with 20 inner-iterations *per* time-step), showing that the method's explicitness (mainly the SIMPLEC algorithm) is the major reason for the error observed. There is no significant difference of accuracy between the results obtained with the coupled and semi-coupled solvers, pointing out to the small effect obtained by using an implicit discretization of the electric force term in the momentum equation, for this particular case.

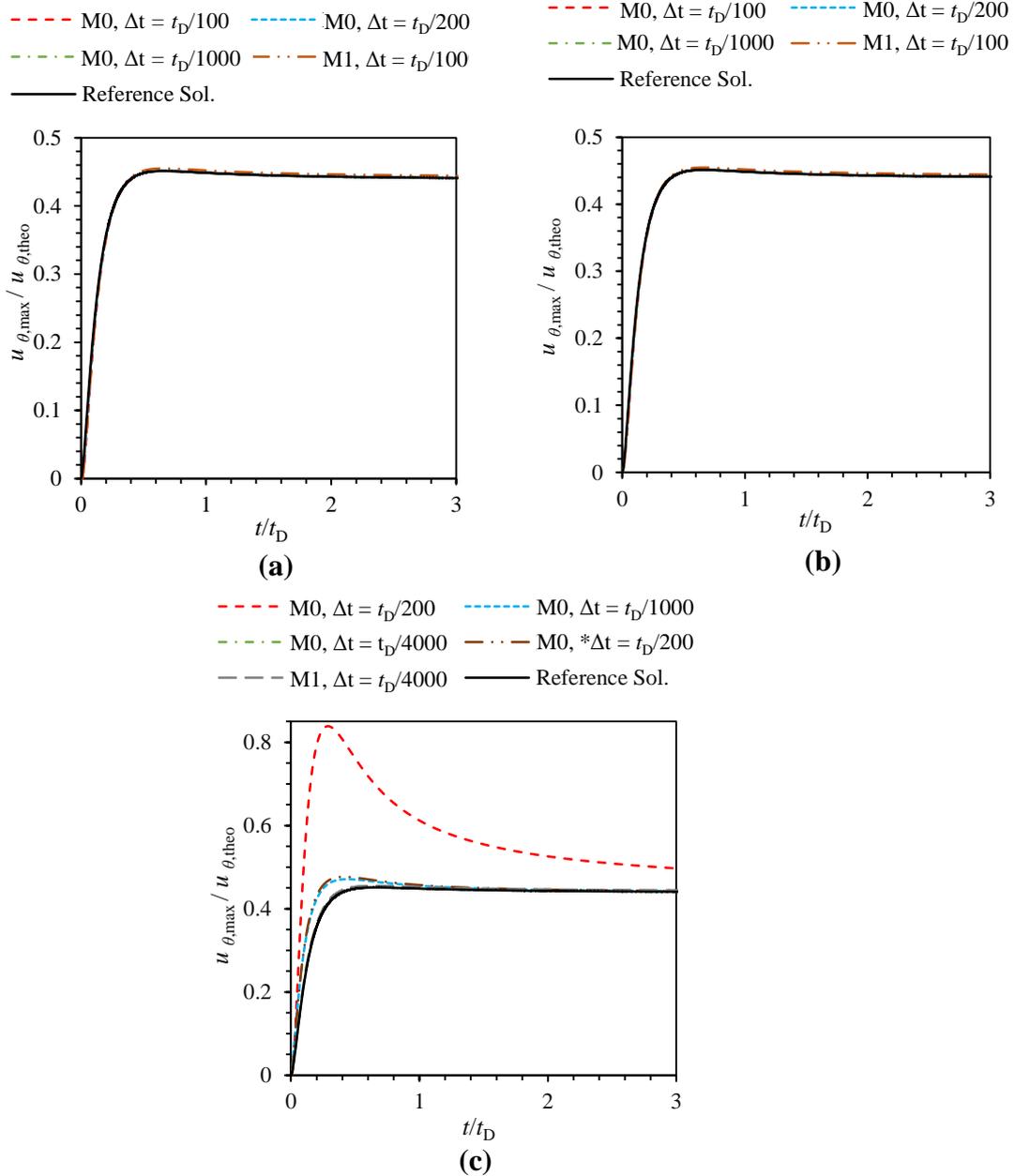

**Figure 5** – Evolution of the maximum azimuthal velocity at $\theta = 3\pi/4$ for $\tilde{\kappa} = 10$ and $\tilde{\zeta} = 1$, using (a) coupled, (b) semi-coupled and (c) segregated solvers. The reference solution plotted in each panel is obtained using a very small time-step. The curve "M0, *$\Delta t = t_D/200$" in panel (c) was obtained using 20 inner-iterations *per* time-step.

The total time of computation for all the set of parameters tested is listed in Table 3. For the same time-step, the coupled and semi-coupled solvers are always slower than the segregated solver. This is no surprise considering the difference of size and complexity of the matrices being solved in each case. However, if accuracy is taken into account, we promptly conclude that segregated solvers are slower. Indeed, taking the ratios for cases with similar levels of accuracy and higher allowable time-steps ($\Delta t = t_D / 100$ for coupled and semi-coupled solvers and $\Delta t = t_D / 4000$ for the segregated solver, Table 3), we observe that the coupled and semi-coupled solvers are 6–11 times faster than the segregated solver. Therefore, although coupled and semi-coupled solvers were significantly slower on a *per* time-step basis, they allowed the use of larger time-steps for the same level of accuracy than the segregated solver, which resulted in a smaller total time of computation. The higher speedup achieved by the semi-coupled solver is a direct consequence of a similar level of accuracy and lower matrix complexity comparing to the coupled solver.

**Table 3** – Computational time for the different solution methods tested in the simulation of the transient ICEO case, for $\tilde{\kappa} = 10$, $\tilde{\zeta} = 1$ and $0 \leq t/t_D \leq 3$.

| Solution method | Matrix solvers | $\Delta t/t_D$ | Total CPU time (s) | | Average CPU time per time-step (s) | |
|---|---|---|---|---|---|---|
| | | | Mesh M0 | Mesh M1 | Mesh M0 | Mesh M1 |
| Coupled | . $p$-**u**-$\Psi$-$c_i$: BiCGStab+LU (reuse) | 1/100 | 320 | 1395 | 1.07 | 4.65 |
| | | 1/200 | 550 | 3034 | 0.92 | 5.06 |
| | | 1/1000 | 2600 | 12021 | 0.87 | 4.01 |
| Semi-coupled | . $p$-**u**: LU (reuse) . $\Psi$-$c_i$: BiCGStab+LU (reuse) | 1/100 | 245 | 965 | 0.82 | 3.22 |
| | | 1/200 | 460 | 2092 | 0.77 | 3.49 |
| | | 1/1000 | 1900 | 7955 | 0.63 | 2.65 |
| Segregated | . $p$, **u**, $\Psi$: GAMG[(*)] . $c_i$: BiCG+DILU[(*)] | 1/200 | 186 | 1426 | 0.31 | 2.38 |
| | | 1/1000 | 697 | 4221 | 0.23 | 1.41 |
| | | 1/4000 | 2163 | 10955 | 0.18 | 0.91 |
| | | 1/200[(†)] | 2338 | 15406 | 3.90 | 25.68 |

(*) Sparse matrix solver from OpenFOAM®.
(†) Using 20 inner iterations per time-step.

The computation times presented in Table 3 were obtained employing the best solving strategy available for each case. However, it is also interesting to compare the performance that would be obtained using different options. Such comparison is presented in Table 4 for some sparse matrix solvers and solving strategies. Among the methods being compared (for similar levels of time accuracy), the semi-coupled solver reusing the preconditioner and factorization is the method with the lowest computational time. It is approximately 11 times faster than the segregated solution method using the GAMG multigrid solver available in OpenFOAM®, which is itself faster (1.2) than a standard preconditioned Krylov solver (CG+DIC) also available in OpenFOAM®. However, the peak memory usage is also increased by a factor of approximately 8. As shown in Table 4, the speedup factor would be significantly smaller (1.57) if the preconditioner and factorization were not reused, even if only direct solvers were applied (1.53). The coupled solver is faster than the segregated solver (7.85), but only if the preconditioner is reused (speedup of 0.9 if not reused). The peak memory usage is the highest (~13 times more than the reference method). Comparing only among the different strategies used with the segregated solver, it can be seen that solving the momentum and continuity equations with a direct method (LU factorization) and further reusing the LU factors allows a speedup factor of 1.61 and requires twice as much memory compared to the reference method. Again, the speedup is only possible by reusing the factorization (speedup of 0.28 without reuse), which, for the segregated solver and this particular test case, only needs to be computed once. Thus, when compared to the default segregated solvers and sparse matrix solvers available in OpenFOAM®, the reduction of the computational time can be achieved not only by using coupled and semi-coupled solvers, but also by using faster, albeit more memory intensive, matrix solvers in the segregated solution method.

**Table 4** – Performance comparison between different sparse matrix solvers for coupled, semi-coupled and segregated solution methods. The results are for mesh M1, $\tilde{\kappa} = 10$, $\tilde{\zeta} = 1$ and $0 \leq t/t_D \leq 3$. The time-step is $\Delta t/t_D = 1/100$ for coupled and semi-coupled solvers, and $\Delta t/t_D = 1/4000$ for segregated solvers in order to ensure a similar level of accuracy among approaches. The peak memory refers to the maximum physical memory usage during the simulation. For segregated solvers, the second column refers to the solver-preconditioner pair used to solve the continuity and momentum equations (the solver for $\Psi$ is GAMG and for $c_i$ is BiCG+DILU). The values inside parentheses in the last two columns represent the ratio relative to the segregated solution method using the GAMG matrix solver from OpenFOAM® (last row).

| Solution method | Sparse matrix solver | Reuse preconditioner/ factorization | Peak memory (Mb) | Total CPU time (s) |
|---|---|---|---|---|
| Coupled | BiCGStab+LU | Yes | 2879 (12.77) | 1395 (7.85) |
| Coupled | BiCGStab+LU | No | 2937 (13.02) | 12128 (0.90) |
| Coupled | LU | No | 2850 (12.64) | 11510 (0.95) |
| Semi-coupled | LU[(1)] and BiCGStab+LU[(2)] | Yes | 1704 (7.55) | 965 (11.35) |
| Semi-coupled | LU[(1)] and BiCGStab+LU[(2)] | No | 899 (3.98) | 6998 (1.57) |
| Semi-coupled | LU[(1,2)] | No | 899 (3.98) | 7179 (1.53) |
| Segregated | LU | Yes | 504 (2.24) | 6800 (1.61) |
| Segregated | LU | No | 323 (1.43) | 39540 (0.28) |
| Segregated | CG+DIC[(*)] | No | 230 (1.02) | 13423 (0.82) |
| Segregated | GAMG[(*)] | No | 226 | 10955 |

(*) Sparse matrix solver from OpenFOAM®.
(1) Sparse matrix solver for the coupled system $p$-**u**.
(2) Sparse matrix solver for the coupled system $\Psi$-$c_i$.

### 5.2.3. Chaotic regime at high voltages

Coupled solvers proved to be advantageous in the steady and transient ICEO cases analyzed. However, their superiority comparing to segregated solvers is probably best evidenced when the imposed voltage is such that $\tilde{\zeta} \gg 1$ and the flow becomes chaotic. In such conditions, the field variables change abruptly in both time and space and a strong coupling between variables is needed to ensure numerical stability. For example, for $\tilde{\zeta} = 50$ (chaotic flow) the ionic concentration changes by more than 5 orders of magnitude within one $\lambda_D$ from the cylinder surface.

We performed simulations for $\tilde{\kappa} = 1000$ and $1 \leq \tilde{\zeta} \leq 50$, keeping the conditions similar to the ones used by Davidson et al. [19], who simulated the chaotic ICEO for the first time. We assume that pressure balances the cylinder surface-normal electric force inside the EDL region. Below $\tilde{\zeta} \approx 30$ the flow converged to a steady solution, whereas above this threshold the flow was chaotic, with all variables fluctuating in time. With the time-step set at $\Delta t = 2 \times 10^{-6} t_D = 2 \frac{\lambda_D^2}{D}$, we have been able to simulate the fluid flow in the whole range of voltages using the coupled and semi-coupled solvers. Moreover, the simulations were also stable if only the PNP system of equations was solved coupled. However, when all the equations were solved segregated, the algorithm diverged for all the range of voltages. The time-step was decreased successively by a factor of 10 (mesh M0) until the (segregated) algorithm became numerically stable. It was shown that a reduction by a factor of 10 was sufficient for $\tilde{\zeta}$ up to ~15, but needed to be reduced by a factor of 100 for $\tilde{\zeta}$ up to ~30. Once the flow became chaotic ($\tilde{\zeta} \gtrsim 30$), not even a time-step reduction by a factor of 100 was enough to avoid numerical divergence. We shall note that this would correspond to $\Delta t = 2 \times 10^{-8} t_D$ and 50 million time-steps would be needed to run the simulation up to $t = t_D$, as was done with the coupled solvers. Even if that time-step value was numerically stable, the total time of computation would render the numerical simulation unfeasible. Therefore, in this case the use of coupled solvers allowed to perform simulations that would be hardly accomplished with segregated solvers.

The results obtained with the coupled solver are plotted in Fig. 6, where we can see a good agreement with the data from Davidson et al. [19]. These simulations were run in parallel to reduce the time of computation. The contours of some relevant quantities are also shown in Fig. 7 for different voltages, in order to show the transition from steady to chaotic flow conditions. The quantities represented are the normalized velocity magnitude ($\tilde{U} = \frac{|\mathbf{u}|}{U} = \frac{|\mathbf{u}|\eta}{\varepsilon_0 \varepsilon_r E^2 b}$), the normalized total ion concentration ($\tilde{c} = \frac{c_+ + c_-}{2c_0}$) and the normalized charge density ($\tilde{\rho}_E = \frac{c_+ - c_-}{c_0}$). The contours of the last two quantities are in qualitative agreement with the patterns shown in Davidson et al. [19], in the chaotic flow regime, where we can observe the formation and migration of ion-depleted plumes (black regions in the contours of $\tilde{c}$) from the poles at $\theta = \pi/2, 3\pi/2$ to the poles at $\theta = 0, \pi$. Moreover, ions ejection at poles $\theta = 0, \pi$ can be clearly observed as a consequence of an enhanced charge transport across the EDL.

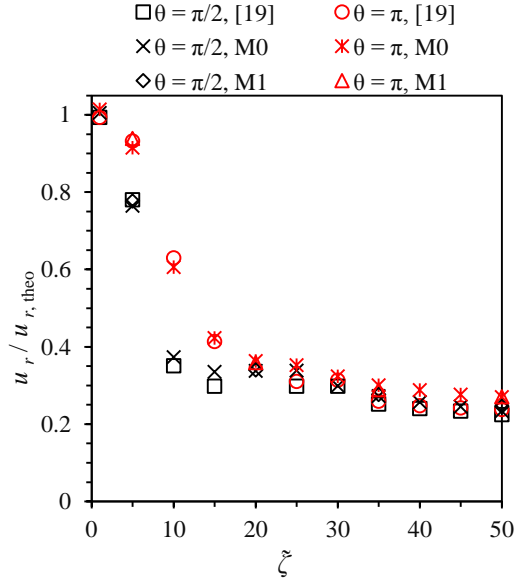

**Figure 6** – Normalized radial velocity component at $\theta = \pi/2$ and $\theta = \pi$, for $\tilde{\kappa} = 1000$ and $r = \sqrt{3}b$. The results obtained with mesh M1 are only plotted for $\tilde{\zeta} = 5, 20, 35$ and $50$. For $\tilde{\zeta} \geq 30$, the points represent time-averaged values, since the flow is chaotic at such voltages.

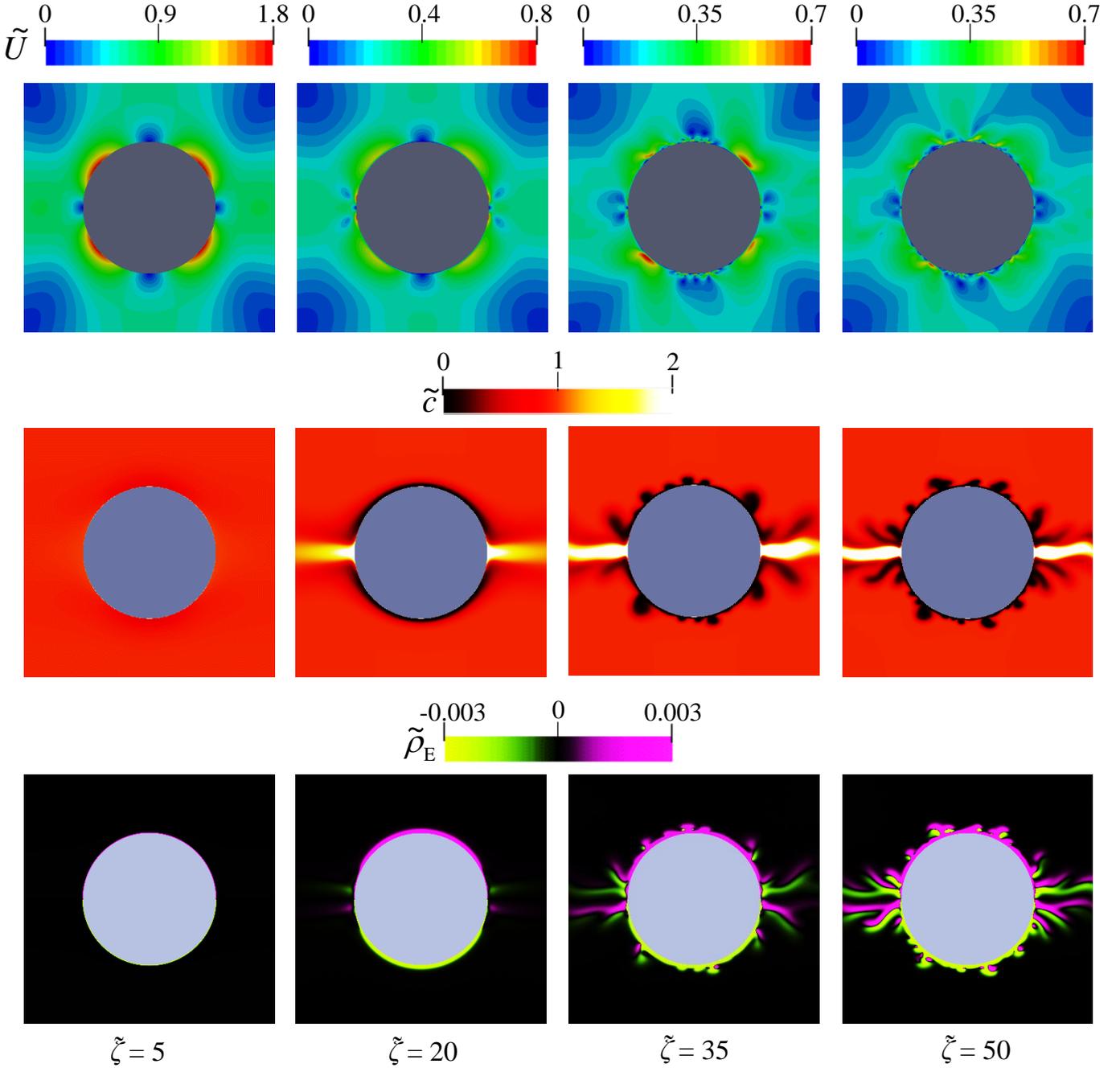

**Figure 7** – Contours of the normalized velocity magnitude, total ion concentration and charge density in the ICEO test case, at different voltages ($\tilde{\zeta} = 5$, 20, 35 and 50, from left to right), for $\tilde{\kappa} = 1000$ and mesh M1. In order to improve visualization, the upper/lower bounds of the scale for the normalized total ion concentration and charge density contours do not correspond to the physical bounds of the given fields. The contours at $\tilde{\zeta} = 35$ and 50 are only representative of the instantaneous patterns at a given instant of time in the interval $0.5 \leq t/t_D \leq 1$, since the flow is chaotic at such voltages.

### 5.3. Case II: electroosmosis in a contraction/expansion device

The second and last test case is the electroosmotic flow in a contraction/expansion device. This flow is numerically challenging due to the singularities developed at the re-entrant corners and it also has experimental relevance [21-23].

The 2D computational domain is depicted in Fig. 8. For the range of parameters tested in this work, the flow is symmetric about plane $y = 0$, thus only half of the geometry was simulated. The half-width of the narrow channel is $H$ and its length is $6H$. The main channel is four times wider and extends up to $200H$ to both sides of the narrow channel. The total channel length is made long enough to reproduce typical experimental conditions. This feature is particularly important in electroosmosis, since both ends of the device are usually kept at the same pressure (pure electroosmosis) and a back-pressure develops over the whole extension of the device due to the fluid flow. This results in a distortion of the characteristic plug-like velocity profile of electroosmotic flows, the extent of which depends on the length of the device. In order to reduce the numerical difficulties that would arise from the use of sharp re-entrant corners, these have been rounded by a small radius of curvature equal to $H/5$.

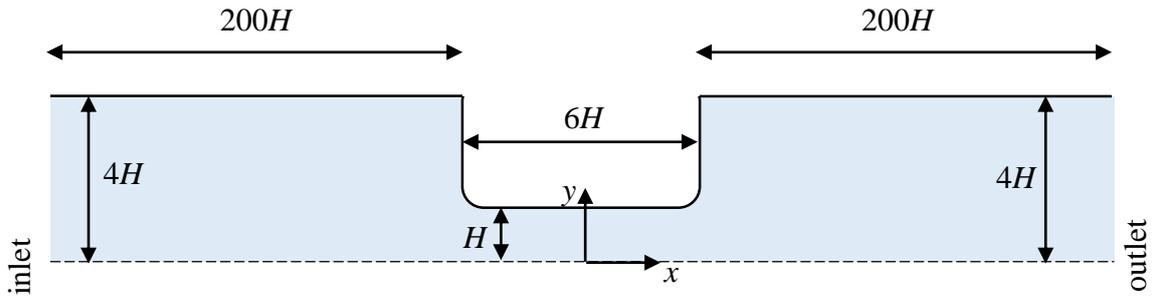

**Figure 8** – Schematic representation of the computational domain for the contraction/expansion case. The domain is composed by three main sections: the entry channel ($8H$ wide and $200H$ long), the contraction channel ($2H$ wide and $6H$ long) and the exit channel ($8H$ wide and $200H$ long). Only half of the domain is represented and simulated due to the plane of symmetry at $y = 0$ (dashed line). The edges at the two re-entrant corners are rounded and the radius of curvature applied is $r = H/5$. The drawing is not scale.

The characteristics of the two grids used to discretize the computational domain are listed in Table 5. Owing to the large dimensions of the geometry and the simultaneous need of a high resolution near the walls while keeping the total number of cells acceptable for simulation, the cells adjacent to the inlet and outlet boundaries have a very high aspect ratio, which poses some challenges to the sparse matrix solvers.

The PB model is employed in this case due to the significantly different time and length scales involved. The following boundary conditions were assigned:

- inlet: $\nabla \psi \cdot \mathbf{n} = 0$, $\varphi = V$, $p = 0$, $\nabla u_i \cdot \mathbf{n} = 0$, $\nabla \tau_{ij} \cdot \mathbf{n} = 0$;
- outlet: $\nabla \psi \cdot \mathbf{n} = 0$, $\varphi = 0$, $p = 0$, $\nabla u_i \cdot \mathbf{n} = 0$, $\nabla \tau_{ij} \cdot \mathbf{n} = 0$;

- walls: $\psi = \psi_w$, $\nabla\varphi \cdot \mathbf{n} = 0$, $\nabla p \cdot \mathbf{n} = 0$, $\mathbf{u} = \mathbf{0}$, $\tau_{ij}$ components are linearly extrapolated from the interior domain [27].

**Table 5** – Details of the two computational meshes used for the contraction/expansion case.

|  | Mesh | |
| --- | --- | --- |
|  | **M0** | **M1** |
| Number of cells in the *y*-direction (main channel) | 100 | 200 |
| Number of cells in the *y*-direction (contraction) | 40 | 80 |
| Total number of cells | 48650 | 194600 |
| Minimum normal cell size adjacent to walls | $\lambda_D/10$ | $\lambda_D/20$ |
| Maximum cell size in the *x*-direction close to the inlet/outlet | 5.6*H* | 2.8*H* |

The simulations were performed considering two ionic species and $H = 100$ μm, $\varepsilon = 0.25$ (extensibility parameter of the PTT model, Eq. 3), $\eta_S = \eta_P = 5\times10^{-4}$ Pa.s, $\rho = 1000$ kg/m$^3$, $V = 50$ V, $\psi_w = -25$ mV, $\varepsilon_r = 84$, $T = 300$ K, $z_+ = -z_- = z = 1$ and $c_0 = 2.5\times10^{-5}$ mol/m$^3$. This set of parameters results in $\tilde{\kappa} = \dfrac{H}{\lambda_D} = 50$. The relaxation time is chosen to impose the intended Deborah number, defined as $De = \dfrac{\lambda U}{H}$, with the velocity scale $U = -\dfrac{\varepsilon_0 \varepsilon_r \psi_w E}{\eta_S + \eta_P}$ and the electric field magnitude $E = \dfrac{4V}{406H}$. The factor of 4 in the numerator of the electric field definition accounts approximately for the contraction effect on the electric field magnitude, thus providing a more realistic value of this quantity inside the contraction channel. For the Newtonian fluid, $De = 0$ and $\eta = \eta_S + \eta_P$. As in the ICEO case, the Reynolds number, defined as $Re = \dfrac{\rho U H}{\eta_S + \eta_P}$, is lower than unity, thus we opted to remove the convective term from the momentum equation, thus imposing creeping flow conditions.

The semi-coupled solver used in this test case only solves the *p*-**u** system of equations coupled. The remaining equations are all solved segregated.

### 5.3.1. Steady-state solution

The steady-state solution was obtained using the different solvers. The system of equations was evolved by time-stepping with a moderate time-step value set at $\Delta t = t_D$, where $t_D = \dfrac{\rho H^2}{\eta_S + \eta_P}$. This strategy allowed to reduce all the residuals below the threshold established for convergence ($10^{-6}$), whereas the pure iterative procedure with under-relaxation failed. Nonetheless, the segregated solution method required an additional under-relaxation of the momentum equation ($\alpha_u = 0.98$) to satisfy the convergence criteria, which is already an indicator of the lower robustness and stability of the segregated solver. As can be seen in Fig. 9(c) and (d), without such additional under-relaxation the residuals stall at a high value for the segregated solver, which is mostly due to the contribution of the problematic high aspect-ratio cells in the mesh.

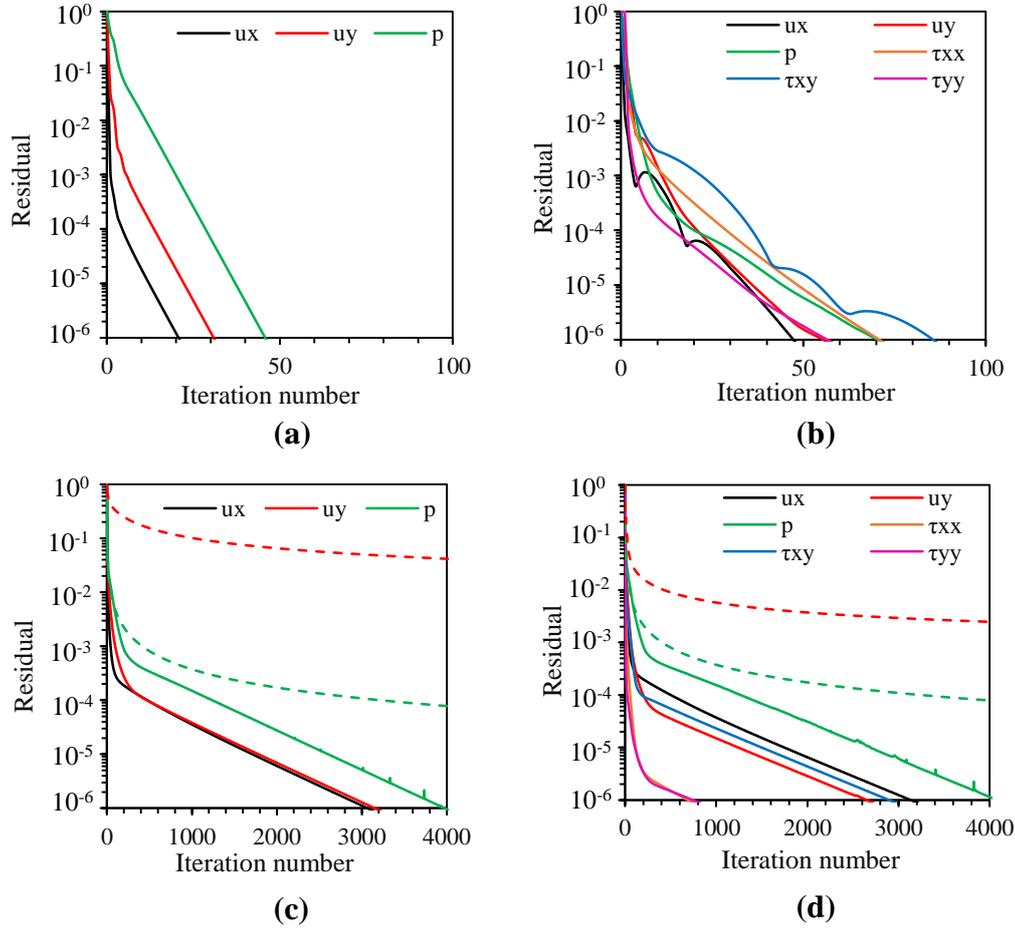

**Figure 9** – Residuals evolution for the contraction/expansion case simulated in mesh M1: (a) $De = 0$, coupled solver; (b) $De = 0.1$, coupled solver; (c) $De = 0$, segregated solver; (d) $De = 0.1$, segregated solver. In panels (c) and (d), the dashed lines represent the residuals for $u_y$ and $p$ when time-stepping is used without under-relaxation.

The performance of each method is presented in Table 6 and the residuals evolution for some of the cases is plotted in Fig. 9. The coupled and semi-coupled solvers always converge in much less iterations and less computational time than the segregated solver. The overall speedup factor ranges from 3 to 99, depending on the conditions and solver. The coupled and semi-coupled solvers converge approximately in the same number of iterations for similar conditions, but the semi-coupled solver always takes less time. This is similar to the behavior observed in the ICEO case.

**Table 6** – Computational time and total number of iterations until convergence for the different solution methods tested in the simulation of the contraction/expansion case. The equations for $\psi$ and $\varphi$ are solved with CG+DIC (from OpenFOAM®) in all cases. The performance data for $De = 0$ in the semi-coupled solution method is the same as for the coupled solution method, since both methods are equivalent for the Newtonian case. The values inside parentheses in the CPU time column represent the speedup factor relative to the segregated solution method using the sparse matrix solvers available in OpenFOAM® (last three rows).

| Solution method | Matrix solvers | $De$ | Total iterations | | Total CPU time (s) | |
|---|---|---|---|---|---|---|
| | | | Mesh M0 | Mesh M1 | Mesh M0 | Mesh M1 |
| Coupled | . $p$-**u**-$\tau$: BiCGStab+LU (reuse) | 0 | 47 | 46 | 31 (17) | 224 (56) |
| | | 0.05 | 60 | 50 | 192 (5) | 869 (20) |
| | | 0.1 | 86 | 86 | 268 (3) | 1420 (13) |
| Semi-coupled | . $p$-**u**: LU (reuse) . $\tau$: BiCG+DILU[(*)] | 0.05 | 67 | 62 | 34 (28) | 180 (99) |
| | | 0.1 | 87 | 87 | 40 (23) | 218 (82) |
| Segregated | . $p$, **u**: LU (reuse) . $\tau$: BiCG+DILU[(*)] | 0 | 1382 | 3975 | 306 (2) | 3698 (3) |
| | | 0.05 | 1391 | 4056 | 461 (2) | 5745 (3) |
| | | 0.1 | 1388 | 4080 | 457 (2) | 5779 (3) |
| Segregated | . $p$, **u**: CG+DIC[(*)] . $\tau$: BiCG+DILU[(*)] | 0 | 1382 | 3975 | 528 | 12564 |
| | | 0.05 | 1391 | 4056 | 925 | 17759 |
| | | 0.1 | 1388 | 4080 | 900 | 17891 |

(*) Sparse matrix solver from OpenFOAM®.

For the coupled and semi-coupled solvers, the number of iterations to convergence does not change significantly when refining the mesh (Table 6), which is an important feature of fully-coupled algorithms [6, 8]. On the other hand, mesh refinement seems to increase considerably the total number of iterations for the segregated solver (by a factor ~3), which increases even more the performance gap to the other two solvers. On the other hand, the effect of $De$ on the number of iterations – convergence rate decreases for increasing $De$ (Table 6) – is apparently more obvious for the coupled and semi-coupled solvers.

In Table 6, we also compare between two strategies that explore the type of sparse matrix solvers used for the momentum and continuity equations in the segregated solver. These equations are typically the ones consuming more time to be solved, with the continuity (pressure) equation being typically the more costly of the two. The first strategy uses a standard sparse matrix solver and preconditioner available in OpenFOAM® (CG+DIC) to solve such equations. The second strategy employs a direct LU solver, reusing the factorization computed in the first time-step (the coefficients of both matrices do not change over time). The residuals evolve equally in the two methods, but the time of computation is different in each case (Table 6). Indeed, the direct solver allows a speedup factor of approximately 2 and 3 in grids M0 and M1, respectively. Although the gain is significantly smaller when compared to the speedup allowed by coupled and semi-coupled solvers, it shows again that the performance of segregated solvers in OpenFOAM® can still be increased in some cases by improving the matrix solving stage. The same has been concluded in [27]. The speedup is expected to be higher for the cases in which the standard iterative solvers available in OpenFOAM® require a high number of iterations to converge to the prescribed tolerance. This happens for matrices with a low condition number, which are frequently related to grids of low quality. In such cases, an iterative sparse matrix solver coupled to a good-quality preconditioner that can be reused, or a direct solver for which the factorization can be reused, will most likely outperform OpenFOAM® matrix solvers in computational speed, but will also use more memory.

The normalized velocity contours obtained for two different $De$ are depicted in Fig. 10 and compared with the Newtonian case ($De = 0$). The distinguishing features for the viscoelastic PTT fluid are the loss of symmetry relative to plane $x = 0$, a higher velocity at the corners, enhanced by the locally large stresses, and a more convex transverse velocity profile inside the contraction. A distinctive characteristic of these EDF compared to the equivalent pressure-driven flows is the absence of vortices which, however, can eventually appear for the PTT fluid at higher $De$, as a result of an elastic instability.

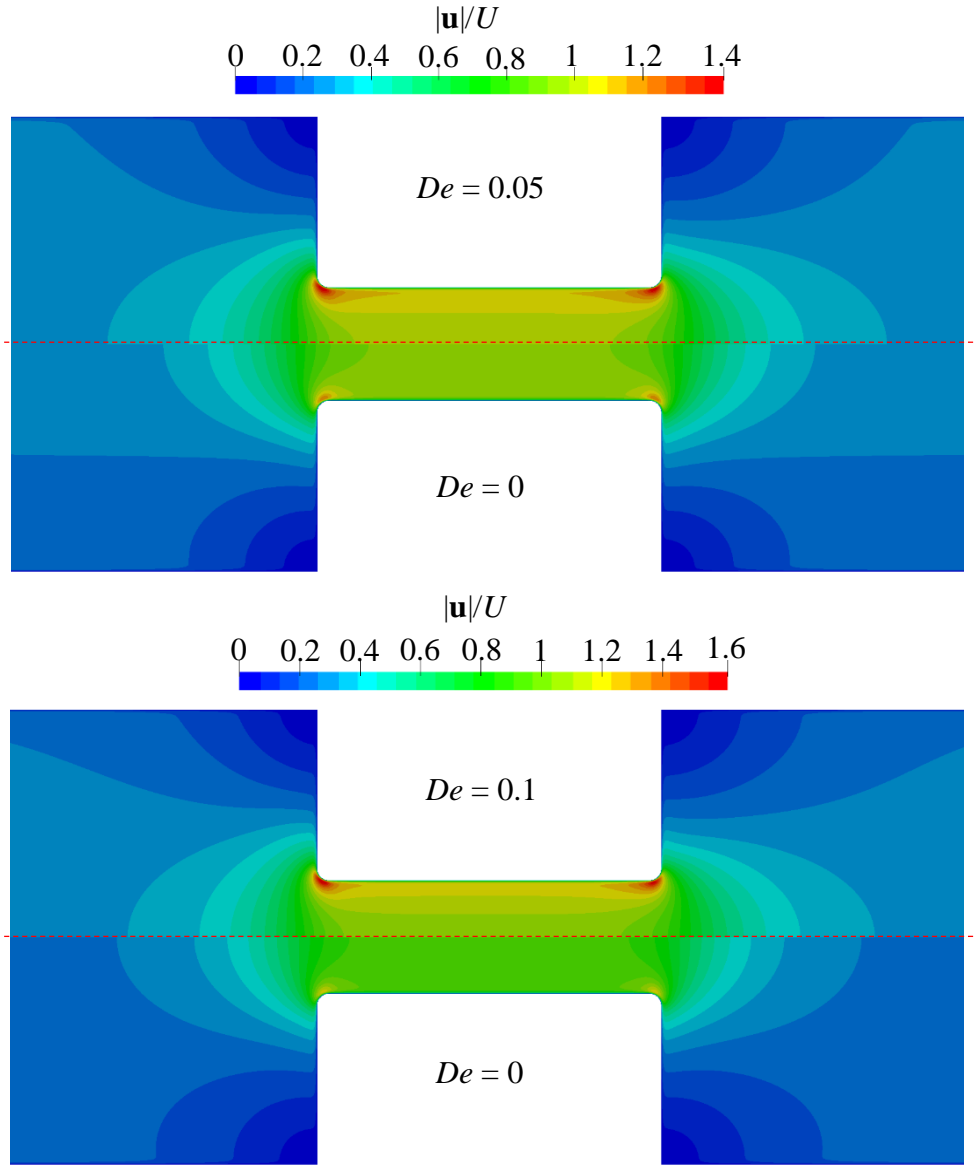

**Figure 10** – Contours of the normalized velocity magnitude in the contraction/expansion case for different *De*. The upper panel compares the contours between *De* = 0 and *De* = 0.05 and the lower panel between *De* = 0 and *De* = 0.1. The flow direction is from left to right.

## 6. Discussion

The coupled and semi-coupled solvers developed in this work were effective in reducing the computational time of steady-state and transient EDF simulations, when compared to the equivalent segregated solvers. This was possible due to essentially three factors. Firstly, the introduction of coupling terms between equations and the assembling of all equations in a single system of equations increased the overall stability of the algorithm, since all variables become implicitly linked. Therefore, larger time-steps and/or higher under-relaxation factors can be used without numerical divergence of the simulations. Secondly, for the same time-step value the coupled solvers are more implicit than segregated solvers and they are more accurate in time, allowing the use of larger time-steps for a similar level of accuracy. The higher level of implicitness arises from the implicit discretization of the coupling terms, some of them replacing explicit or semi-implicit coupling algorithms used in the segregated solver, as the SIMPLEC algorithm for pressure-velocity coupling. The third and last factor was the use of efficient sparse matrix solvers, including the reuse of preconditioners and factorizations whenever possible.

The scalability of the solvers with mesh size was not explored in detail (cases were always limited to only two different grid sizes), but the results obtained in the two test cases suggest that both coupled and semi-coupled solvers are able to keep the convergence rate independent from the grid size in most of the situations analyzed. In such conditions, the scaling of the total computational time with grid size is expected to depend mainly on the scalability of the matrix solvers (limiting step of the overall algorithm). On the other hand, the convergence rate of segregated solvers showed a significant dependence on grid size.

In all the cases tested, semi-coupled solvers showed better performance than coupled solvers, presenting similar accuracy, lower computational time and lower memory usage. Therefore, dropping some implicit coupling terms seems to be advantageous. In most of the cases tested in this work, the pressure-velocity coupling was seen to have a higher importance than the coupling between other variables. The exception was the ICEO test case in chaotic flow conditions, where the coupling between the electric potential and species concentration was key to avoid numerical divergence. The relative importance between the several coupling relations is case-dependent and should be taken into account while deciding about the best solver for a given case. Due to the much lower computational cost per iteration of segregated solvers, the implicit coupling between a certain pair of variables should only be used if there are gains in stability or, indirectly, in the overall computational time. Moreover, for the viscoelastic case in particular, a semi-coupled solver where the polymeric extra-stresses are solved segregated (no implicit coupling between extra-stresses and velocity) would allow applying popular stabilization techniques relying on changes of variable, such as the log-conformation tensor approach, that could not be used in a coupled solver.

The matrix solving stage is at the core of a coupled/semi-coupled solution method and, in our view, represents the current challenge of these methods. It is of little interest building a coupled system of equations if it cannot be solved efficiently. In this work, we relied mostly on direct solvers and LU preconditioned iterative solvers. These solvers proved to be cost-effective as long as the factorization/preconditioner can be reused during the simulation. In some cases, the factorization/preconditioner was computed once, at the beginning of the computations, whereas in other cases it was re-computed every *n* time-steps. This recycling strategy is not expected to be effective in all cases, but it can be successfully applied in a number of CFD simulations. For example, in microfluidics it is frequently possible to neglect momentum advection and, in such situations, the matrix of coefficients for the momentum and pressure equations does not change over time for a fixed time-step and viscosity, which allows reusing the preconditioner/factorization for the respective matrices. Moreover, this strategy can be also applied to segregated solvers, as shown in Sections 5.2.2 and 5.3.1, where the time of computation dropped by a factor of 3. Nonetheless, direct solvers and complete LU preconditioning are memory intensive methods – this is the price to pay for robustness. Therefore, its applicability is also limited by the size of the computational domain, although it is worth noting that there is also practical interest in simulations with a small-medium number of cells. In addition to memory and CPU time, an equally important indicator of performance is the scalability of the methods in parallel computations. This subject was not addressed in this work, because optimization is underway to improve scalability. Note, however, that all methods were implemented in parallel and can use an arbitrary number of processors (the results for chaotic ICEO were obtained using parallel simulations). The future research in this area should encompass efficient iterative sparse matrix solvers (e.g. [13]) owing to their low memory usage.

## 7. Conclusions

This work addressed the implementation of coupled solvers in OpenFOAM® to simulate transient and steady-state electrically-driven flows. The coupling involves the continuity equation, momentum balance, the constitutive equation for extra-stresses (in case of viscoelastic fluids) and the Poisson-Nernst-Planck system of equations for ions transport, which can be replaced by the Poisson-Boltzmann system of equations. The method presented is generic in scope and valid for Newtonian fluids and a number of viscoelastic fluid models, although only Newtonian and linear PTT fluids were analyzed in this work. The resulting coupled systems of equations were solved with PETSc library through a parallelized interface specifically built for that purpose.

The coupled solvers compared successfully against an analytical solution and their performance was assessed and compared with segregated solvers in two test cases, including the induced-charge electroosmosis of a Newtonian fluid around a cylinder and the electroosmotic flow of a linear PTT fluid in a contraction/expansion microchannel. In transient flow problems, the coupled solvers allowed to use larger time-steps for a similar level of accuracy, thus requiring less iterations to simulate a given period of physical time. This superiority was best evidenced in the simulation of chaotic induced-charge electroosmosis, since the segregated solver simply failed to retrieve a solution in a feasible amount of time, whereas the results obtained with the coupled solver agreed well with the reference data. For steady-state flow problems, larger time-steps and/or higher under-relaxation factors could be used with coupled solvers, thus converging in fewer iterations. Overall, the coupled solvers produced speedups ranging from 3 to 99. The highest speedup factors were obtained with semi-coupled solvers, which drop the implicit coupling between some pairs of variables. These semi-coupled solvers presented the best performance (same accuracy, lower computational time and lower memory usage compared to coupled solvers) in all the test cases addressed in this work.

The coupled systems of equations assembled in this work were mostly solved using either direct solvers or complete LU factorization as preconditioner. While reusing the factorization or preconditioner proved to be an effective way to speed up the computations, the memory requirements of such methods was relatively high and the recycling policy has a limited scope of application. Therefore, the continuation of this work, and more generally the research in the area of coupled solution methods, should essentially address the optimization of the matrix solving stage, which is currently the bottleneck of the method. This includes not only the development of more efficient and more robust sparse matrix solvers and preconditioners, but also their efficient and scalable implementation in central and graphical processing units.

The solvers developed in this work were incorporated in rheoTool [20], being available in open-source. Although this study focused specifically on electrically-driven flows, some modules of the code are general enough to be applied to the coupling between equations from other areas of CFD. Moreover, the interfaces to external open-source libraries (PETSc, Hypre and Eigen), developed during the course of this work, allow solving any generic matrix generated in OpenFOAM® with the sparse matrix solvers available in such libraries.

**References**

[1] A.J. Chorin, Numerical solution of the Navier-Stokes equations, Mathematics of computation, 22 (1968) 745-762, https://doi.org/10.2307/2004575.

[2] F. Moukalled, L. Mangani, M. Darwish, The finite volume method in computational fluid dynamics: an advanced introduction with OpenFOAM and Matlab, Springer Publishing Company, Incorporated, 2015.

[3] F.S. Sousa, C.M. Oishi, G.C. Buscaglia, Spurious transients of projection methods in microflow simulations, Computer Methods in Applied Mechanics and Engineering, 285 (2015) 659-693, https://doi.org/10.1016/j.cma.2014.11.039.

[4] S.C. Xue, R.I. Tanner, N. Phan-Thien, Numerical modelling of transient viscoelastic flows, Journal of Non-Newtonian Fluid Mechanics, 123 (2004) 33-58, https://doi.org/10.1016/j.jnnfm.2004.06.009.

[5] M. Darwish, D. Asmar, F. Moukalled, A comparative assessment within a multigrid environment of segregated pressure-based algorithms for fluid flow at all speeds, Numerical Heat Transfer, Part B: Fundamentals, 45 (2004) 49-74, https://doi.org/10.1080/1040779049025487.

[6] L. Mangani, M. Buchmayr, M. Darwish, Development of a novel fully coupled solver in OpenFOAM: steady-state incompressible turbulent flows, Numerical Heat Transfer, Part B: Fundamentals, 66 (2014) 1-20, https://doi.org/10.1080/10407790.2014.894448.

[7] P.M. Gresho, Incompressible fluid dynamics: some fundamental formulation issues, Annual Review of Fluid Mechanics, 23 (1991) 413-453, https://doi.org/10.1146/annurev.fl.23.010191.002213.

[8] M. Darwish, I. Sraj, F. Moukalled, A coupled finite volume solver for the solution of incompressible flows on unstructured grids, Journal of Computational Physics, 228 (2009) 180-201, https://doi.org/10.1016/j.jcp.2008.08.027.

[9] L. Mangani, M. Darwish, F. Moukalled, An OpenFOAM pressure-based coupled CFD solver for turbulent and compressible flows in turbomachinery applications, Numerical Heat Transfer, Part B: Fundamentals, 69 (2016) 413-431, https://doi.org/10.1080/10407790.2015.1125212.

[10] C. Fernandes, V. Vukčević, T. Uroić, R. Simoes, O.S. Carneiro, H. Jasak, J.M. Nóbrega, A coupled finite volume flow solver for the solution of incompressible viscoelastic flows, Journal of Non-Newtonian Fluid Mechanics, 265 (2019) 99-115, https://doi.org/10.1016/j.jnnfm.2019.01.006.

[11] C.-N. Xiao, F. Denner, B.G.M. van Wachem, Fully-coupled pressure-based finite-volume framework for the simulation of fluid flows at all speeds in complex geometries, Journal of Computational Physics, 346 (2017) 91-130, https://doi.org/10.1016/j.jcp.2017.06.009.

[12] F. Denner, Fully-coupled pressure-based algorithm for compressible flows: Linearisation and iterative solution strategies, Computers & Fluids, 175 (2018) 53-65, https://doi.org/10.1016/j.compfluid.2018.07.005.

[13] T. Uroić, H. Jasak, Block-selective algebraic multigrid for implicitly coupled pressure-velocity system, Computers & Fluids, 167 (2018) 100-110, https://doi.org/10.1016/j.compfluid.2018.02.034.

[14] E. Karatay, C.L. Druzgalski, A. Mani, Simulation of chaotic electrokinetic transport: Performance of commercial software versus custom-built direct numerical simulation codes, Journal of Colloid and Interface Science, 446 (2015) 67-76, https://doi.org/10.1016/j.jcis.2014.12.081.

[15] F. Pimenta, M.A. Alves, Numerical simulation of electrically-driven flows using OpenFOAM, arXiv:1802.02843, (2018) https://arxiv.org/abs/1802.02843.

[16] V.S. Pham, Z. Li, K.M. Lim, J.K. White, J. Han, Direct numerical simulation of electroconvective instability and hysteretic current-voltage response of a permselective membrane, Physical Review E, 86 (2012) 046310, https://doi.org/10.1103/PhysRevE.86.046310.
[17] E.A. Demekhin, V.S. Shelistov, S.V. Polyanskikh, Linear and nonlinear evolution and diffusion layer selection in electrokinetic instability, Physical Review E, 84 (2011) 036318, https://doi.org/10.1103/PhysRevE.84.036318.
[18] J.D. Sherwood, M. Mao, S. Ghosal, Electroosmosis in a finite cylindrical pore: simple models of end effects, Langmuir, 30 (2014) 9261-9272, https://doi.org/10.1021/la502349g.
[19] S.M. Davidson, M.B. Andersen, A. Mani, Chaotic induced-charge electro-osmosis, Physical Review Letters, 112 (2014) 128302, https://doi.org/10.1103/PhysRevLett.112.128302.
[20] F. Pimenta, M.A. Alves, rheoTool, https://github.com/fppimenta/rheoTool.
[21] R.M. Bryce, M.R. Freeman, Abatement of mixing in shear-free elongationally unstable viscoelastic microflows, Lab on a Chip, 10 (2010) 1436-1441, https://doi.org/10.1039/B925391B.
[22] R.M. Bryce, M.R. Freeman, Extensional instability in electro-osmotic microflows of polymer solutions, Physical Review E, 81 (2010) 036328, https://doi.org/10.1103/PhysRevE.81.036328.
[23] C.-H. Ko, D. Li, A. Malekanfard, Y.-N. Wang, L.-M. Fu, X. Xuan, Electroosmotic flow of non-Newtonian fluids in a constriction microchannel, Electrophoresis, 0 (2018) https://doi.org/10.1002/elps.201800315.
[24] F. Pimenta, M.A. Alves, Electro-elastic instabilities in cross-shaped microchannels, Journal of Non-Newtonian Fluid Mechanics, 259 (2018) 61-77, https://doi.org/10.1016/j.jnnfm.2018.04.004.
[25] H.M. Park, J.S. Lee, T.W. Kim, Comparison of the Nernst–Planck model and the Poisson–Boltzmann model for electroosmotic flows in microchannels, Journal of Colloid and Interface Science, 315 (2007) 731-739, https://doi.org/10.1016/j.jcis.2007.07.007.
[26] A. Prohl, M. Schmuck, Convergent finite element discretizations of the Navier-Stokes-Nernst-Planck-Poisson system, ESAIM: M2AN, 44 (2010) 531-571, https://doi.org/10.1051/m2an/2010013.
[27] F. Pimenta, M.A. Alves, Stabilization of an open-source finite-volume solver for viscoelastic fluid flows, Journal of Non-Newtonian Fluid Mechanics, 239 (2017) 85-104, https://doi.org/10.1016/j.jnnfm.2016.12.002.
[28] R. Guénette, M. Fortin, A new mixed finite element method for computing viscoelastic flows, Journal of Non-Newtonian Fluid Mechanics, 60 (1995) 27-52, https://doi.org/10.1016/0377-0257(95)01372-3.
[29] C. Fernandes, M.S.B. Araujo, L.L. Ferrás, J. Miguel Nóbrega, Improved both sides diffusion (iBSD): A new and straightforward stabilization approach for viscoelastic fluid flows, Journal of Non-Newtonian Fluid Mechanics, 249 (2017) 63-78, https://doi.org/10.1016/j.jnnfm.2017.09.008.
[30] N.P. Thien, R.I. Tanner, A new constitutive equation derived from network theory, Journal of Non-Newtonian Fluid Mechanics, 2 (1977) 353-365, https://doi.org/10.1016/0377-0257(77)80021-9.
[31] B.J. Kirby, Micro- and nanoscale fluid mechanics: transport in microfluidic devices, Cambridge University Press, 2010.
[32] R. Fattal, R. Kupferman, Constitutive laws for the matrix-logarithm of the conformation tensor, Journal of Non-Newtonian Fluid Mechanics, 123 (2004) 281-285, https://doi.org/10.1016/j.jnnfm.2004.08.008.

[33] J.P. Van Doormaal, G.D. Raithby, Enhancements of the SIMPLE method for predicting incompressible fluid flows, Numerical Heat Transfer, 7 (1984) 147-163, https://doi.org/10.1080/01495728408961817.

[34] M.A. Alves, P.J. Oliveira, F.T. Pinho, A convergent and universally bounded interpolation scheme for the treatment of advection, International Journal for Numerical Methods in Fluids, 41 (2003) 47-75, https://doi.org/10.1002/fld.428.

[35] J. Fish, T. Belytschko, A first course in finite elements, John Wiley & Sons, New York, 2007.

[36] Y. Saad, Iterative methods for sparse linear systems, Society for Industrial and Applied Mathematics, 2003.

[37] H. van der Vorst, Bi-CGSTAB: a fast and smoothly converging variant of Bi-CG for the solution of nonsymmetric linear systems, SIAM Journal on Scientific and Statistical Computing, 13 (1992) 631-644, https://doi.org/10.1137/0913035.

[38] B. Jacob, G. Guennebaud, Eigen, http://eigen.tuxfamily.org.

[39] HYPRE: High performance preconditioners, http://www.llnl.gov/CASC/hypre/.

[40] S. Balay, S. Abhyankar, M.F. Adams, J. Brown, P. Brune, K. Buschelman, L. Dalcin, A. Dener, V. Eijkhout, W.D. Gropp, D. Kaushik, M.G. Knepley, D.A. May, L.C. McInnes, R.T. Mills, T. Munson, K. Rupp, P. Sanan, B.F. Smith, S. Zampini, H. Zhang, H. Zhang, PETSc Users Manual, in, Argonne National Laboratory, 2018.

[41] S. Balay, S. Abhyankar, M.F. Adams, J. Brown, P. Brune, K. Buschelman, L. Dalcin, A. Dener, V. Eijkhout, W.D. Gropp, D. Kaushik, M.G. Knepley, D.A. May, L.C. McInnes, R.T. Mills, T. Munson, K. Rupp, P. Sanan, B.F. Smith, S. Zampini, H. Zhang, H. Zhang, PETSc Web page, in, 2018.

[42] S. Balay, W.D. Gropp, L.C. McInnes, B.F. Smith, Efficient Management of Parallelism in Object Oriented Numerical Software Libraries, in: Modern Software Tools in Scientific Computing, Birkhäuser Press, 1997, pp. 163-202.

[43] P. Amestoy, I. Duff, J. L'Excellent, J. Koster, A fully asynchronous multifrontal solver using distributed dynamic scheduling, SIAM Journal on Matrix Analysis and Applications, 23 (2001) 15-41, https://doi.org/10.1137/S0895479899358194.

[44] P.R. Amestoy, A. Guermouche, J.-Y. L'Excellent, S. Pralet, Hybrid scheduling for the parallel solution of linear systems, Parallel Computing, 32 (2006) 136-156, https://doi.org/10.1016/j.parco.2005.07.004.

[45] A.M. Afonso, M.A. Alves, F.T. Pinho, Analytical solution of mixed electro-osmotic/pressure driven flows of viscoelastic fluids in microchannels, Journal of Non-Newtonian Fluid Mechanics, 159 (2009) 50-63, https://doi.org/10.1016/j.jnnfm.2009.01.006.

[46] T.M. Squires, M.Z. Bazant, Induced-charge electro-osmosis, Journal of Fluid Mechanics, 509 (2004) 217-252, https://doi.org/10.1017/S0022112004009309.